\documentclass[pdflatex,sn-mathphys-num]{sn-jnl}

\usepackage{graphicx}%
\usepackage{multirow}%
\usepackage{amsmath,amssymb,amsfonts}%
\usepackage{amsthm}%
\usepackage{mathrsfs}%
\usepackage[title]{appendix}%
\usepackage{xcolor}%
\usepackage{textcomp}%
\usepackage{manyfoot}%
\usepackage{booktabs}%
\usepackage{algorithm}%
\usepackage{algorithmicx}%
\usepackage{algpseudocode}%
\usepackage{listings}%

\theoremstyle{thmstyleone}%
\theoremstyle{thmstyletwo}%

\theoremstyle{thmstylethree}%

\begin{document}

\title[Article Title]{Strain-tunable inter-valley scattering defines universal mobility enhancement in n- and p-type 2D TMDs}


\author[1,5,6]{\fnm{Sheikh Mohd Ta-Seen} \sur{Afrid}}\email{safrid2@illinois.edu}

\author[1,5,6]{\fnm{He Lin} \sur{Zhao}}\email{helinlz2@illinois.edu}

\author[2,3,4,5,6]{\fnm{Arend M. van der} \sur{Zande}}\email{arendv@illinois.edu}

\author*[1,5,6]{\fnm{Shaloo} \sur{Rakheja}}\email{rakheja@illinois.edu}

\affil[1]{\orgdiv{Department of Electrical and Computer Engineering}, \orgname{University of Illinois Urbana-Champaign}, \orgaddress{\city{Urbana}, \postcode{61801}, \state{Illinois}, \country{United States}}}

\affil[2]{\orgdiv{Department of Materials Science and Engineering}, \orgname{University of Illinois Urbana-Champaign}, \orgaddress{\city{Urbana}, \postcode{61801}, \state{Illinois}, \country{United States}}}

\affil[3]{\orgdiv{Department of Mechanical Science and Engineering}, \orgname{University of Illinois Urbana-Champaign}, \orgaddress{\city{Urbana}, \postcode{61801}, \state{Illinois}, \country{United States}}}

\affil[4]{\orgdiv{Materials Research Laboratory}, \orgname{University of Illinois Urbana-Champaign}, \orgaddress{\city{Urbana}, \postcode{61801}, \state{Illinois}, \country{United States}}}

\affil[5]{\orgdiv{Holonyak Micro and Nanotechnology Laboratory}, \orgname{University of Illinois Urbana-Champaign}, \orgaddress{\city{Urbana}, \postcode{61801}, \state{Illinois}, \country{United States}}}

\affil[6]{\orgdiv{Grainger College of Engineering}, \orgname{University of Illinois Urbana-Champaign}, \orgaddress{\city{Urbana}, \postcode{61801}, \state{Illinois}, \country{United States}}}


\abstract{

Strain fundamentally alters carrier transport in semiconductors by modifying their band structure and scattering pathways. In transition-metal dichalcogenides (TMDs), an emerging class of 2D semiconductors, we show that mobility modulation under biaxial strain is dictated by changes in inter-valley scattering rather than effective mass renormalization as in bulk silicon. Using a multiscale full-band transport framework that incorporates both intrinsic phonon, extrinsic impurity, and dielectric scattering, we find that tensile strain enhances n-type mobility through K--Q valley separation, while compressive strain improves p-type mobility via $\Gamma$--K decoupling. The tuning rates calculated from our full-band model far exceed those achieved by strain engineering in silicon. Both relaxed and strain-modulated carrier mobilities align quantitatively with experimentally verified measurements and are valid across a wide range of practical FET configurations. The enhancement remains robust across variations in temperature, carrier density, impurity level, and dielectric environment. Our results highlight the pivotal role of strain in improving the reliability and performance of 2D TMD-based electronics.
}


\maketitle

\section*{Introduction}
\label{sec:introduction}

Strain engineering has proven to be a powerful technique for enhancing the electronic properties of semiconductors over the past decades \citep{santra2024strain, jaikissoon2024cmos, chaves2020bandgap}. As transistor nodes scale into the nanosheet regime, the nanometer-thin channels are becoming increasingly susceptible to parasitic strain effects, necessitating proper strain engineering to ensure optimal performance. 
Simultaneously, 2D materials, specifically transition metal dichalcogenides (TMDs), have emerged as highly desirable channel material candidates for beyond-silicon nanosheet field-effect transistors (FETs) \citep{kanungo20222d, cao2021dissipative}. Unlike bulk semiconductors, atomically thin TMDs can withstand remarkable levels of elastic deformation without fracture, typically up to 6-11\% strain \citep{liu2025reduction, basu2023strain, duerloo2012intrinsic, bertolazzi2011stretching}. This mechanical resilience enables the nanosheet channels to withstand harsher fabrication processes and unlocks great potential for performance enhancement through strain engineering. 

TMDs such as MoS\textsubscript{2}, MoSe\textsubscript{2}, and WS\textsubscript{2}  exhibit n-type behavior, characterized by electrons as majority carriers \citep{zhang2024enhancing, datye2022strain, zhang2021rapid, yang2024biaxial, wang2021electron}. In contrast, p-type TMDs, such as WSe\textsubscript{2}, MoTe\textsubscript{2}, and in some cases MoSe\textsubscript{2} (depending on doping or substrate interaction), exhibit hole-dominated conduction \citep{nutting2021electrical, chen2017highly, ghosh2025high, bae2021mote2, shang2020situ}. The performance of FETs utilizing TMDs as channel materials is fundamentally determined by the efficiency of charge carrier transport, typically quantified by carrier mobility. High carrier mobility leads to faster switching speeds and lower power consumption in electronic devices \citep{ng2022improving}. In ideal, defect-free crystals, the maximum achievable mobility is limited by interactions between charge carriers and lattice vibrations \citep{cheng2018limits}. However, in practical device settings, other factors often dominate and significantly reduce carrier mobility below this theoretical limit \citep{das2021transistors}. Extrinsic mobility-limiting factors include Coulomb scattering from charged impurities in the substrate or within the material itself \citep{li2016charge} and remote phonon scattering from polar substrate materials \citep{mansoori2023mobility}. Understanding and controlling these mobility-limiting mechanisms is crucial for advancing 2D material-based technology.

Applying strain to TMD monolayers can modify their electronic band structure~\citep{katiyar2025strain, moghal2022tuning, frisenda2017biaxial}, and alter the energy separation between different valleys in the band structure~\citep{sahu2024strain, zhou2025engineering}. Also, strain affects electron-phonon interactions by altering phonon frequencies and modifying the deformation potentials that determine carrier-phonon coupling strength \citep{pan2024strain}. Several experimental and theoretical studies have demonstrated that tensile strain can enhance carrier mobility in specific TMDs \citep{datye2022strain, chen2021carrier, yu2015phase}. In comparison, the role of compressive strain remains less understood. Early findings suggest that it can significantly modify the band structure and induce energy separation between different valleys, providing a distinct yet potentially valuable form of electronic tunability \citep{islam2024strain, maniadaki2016strain}. 
To paint a comprehensive picture of the effect that strain has on TMD transport, both mobility-enhancing and degrading effects caused by both tensile and compressive strain must be studied.
 
Previous strain engineering studies have predominantly focused on a single material, typically MoS\textsubscript{2} \citep{zhang2024enhancing, chen2023mobility, datye2022strain, chen2021carrier, yang2024biaxial} or WS\textsubscript{2} \citep{yang2024biaxial}, and have primarily examined electron transport, with limited attention given to strain-modulated hole transport. In this study, we extend these datasets by performing computations on the strain tuning of hole transport in addition to electron transport. More critically, we go beyond the effective mass approximation used in earlier studies \citep{phuc2018tuning, yu2015phase, sun2018first, cheng2018limits, hosseini2015strainmos2, hosseini2015strain, li2013intrinsic, jin2014intrinsic, kumar2024strainfet}, which oversimplifies the complex band structure of TMDs. As such, our method more accurately captures important phenomena such as non-parabolic bands and multi-valley effects, which are crucial for properly modeling inter-valley scattering processes \citep{pimenta2022electronic}.

Also notably, we include the previously neglected interplay between strain-modified intrinsic scattering and extrinsic scattering mechanisms that are inevitably present in real devices \citep{yu2015phase, phuc2018tuning, cheng2018limits}. Charged impurity scattering and surface optical phonon scattering from substrates typically dominate mobility in practical transistor nodes \citep{li2016charge, mansoori2023mobility, rosul2022effect}, and we extend this consideration to 2D devices. Since the rate of mobility enhancement likely depends on material-specific properties including strain level, temperature, carrier concentration, impurity density, and the dielectric environment, our study connects key experimental parameters to the underlying physics of competing scattering mechanisms in both n- and p-type TMDs.

We establish a unified physical scheme to enable the rational design of high-performance strain-engineered devices by developing a multiscale modeling framework that integrates first-principles calculations with a full-band transport model. 
Our approach simultaneously accounts for intrinsic and extrinsic scattering mechanisms, including acoustic deformation potential (ADP), optical deformation potential (ODP), polar optical phonon (POP), inter-valley (IV), piezoelectric (PZ), charged impurity (CI), and surface optical phonon (SOP) scatterings from various dielectric environments. We focus on the effects of biaxial strain, which produces uniform expansion or compression in the material plane and is particularly effective for modifying electronic band structures. We apply the full-band transport model to three n-type TMDs (MoS\textsubscript{2}, MoSe\textsubscript{2}, and WS\textsubscript{2}) and three p-type TMDs (MoSe\textsubscript{2}, WSe\textsubscript{2}, and MoTe\textsubscript{2}) under biaxial strain conditions. Our calculations span a wide range of experimentally accessible parameters, enabling direct comparison with observations. We quantitatively mapped how the mobility enhancement rate varies across the multi-dimensional parameter space. This provides a practical point of reference for tailoring and optimizing the performance of strain-engineered 2D TMD-based applications, from high-speed transistors to more-than-Moore applications such as flexible electronics.

\section*{Results}\label{sec2}

\subsection*{Electron mobility enhancement}

\begin{figure}[htp]
\centering
\includegraphics[width=0.97\textwidth]{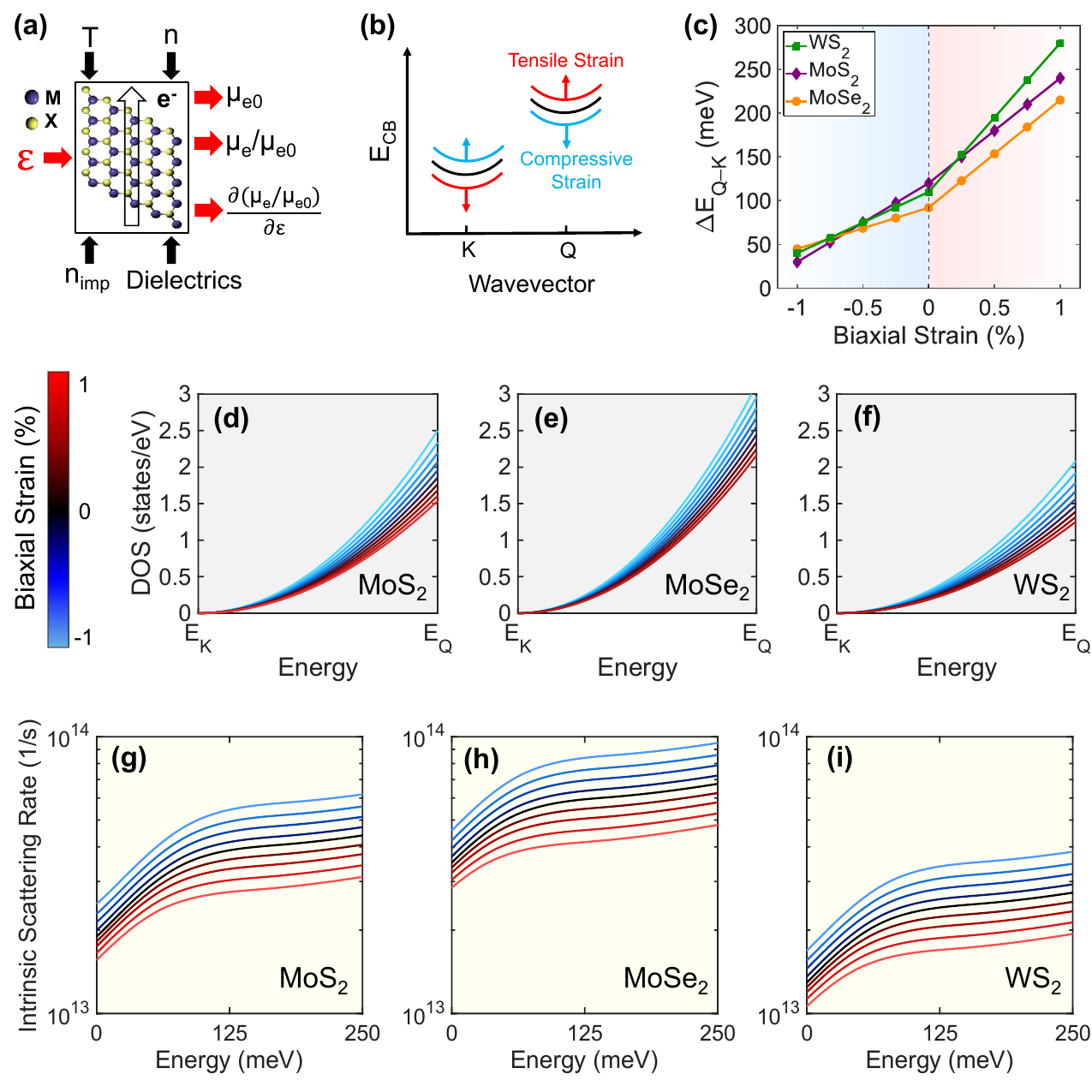}
\caption{\textbf{Mechanisms of strain-induced electron mobility enhancement in n-type TMDs.} 
\textbf{a,} Schematic illustration of the key parameters governing electron mobility in n-type TMDs, highlighting the functional relationship between mobility enhancement and applied strain. 
\textbf{b,} Schematic of the conduction band structure showing energy separation between K and Q valleys ($\Delta \mathrm{E}_{\mathrm{QK}}$) under biaxial strain. 
\textbf{c,} Evolution of the energy separation $\Delta {\mathrm{E_{QK}}}$ under biaxial strain for MoS$_2$ (purple), MoSe$_2$ (orange), and WS$_2$ (green). First-principles calculated density of states versus conduction band energy along the K–Q direction under varying strain for \textbf{d,} MoS$_2$, \textbf{e,} MoSe$_2$, and \textbf{f,} WS$_2$. Computed intrinsic scattering rates versus carrier energy under varying strain for \textbf{g,} MoS$_2$, \textbf{h,} MoSe$_2$, and \textbf{i,} WS$_2$, showing suppression of scattering under tensile strain. In all panels, compressive, unstrained, and tensile strain regimes are represented by blue, black, and red, respectively.}
\label{fig:ntype}
\end{figure}

Here we present a comprehensive investigation of electron mobility enhancement in monolayer n-type TMDs--specifically MoS$_2$, MoSe$_2$, and WS$_2$, under applied biaxial strain. The relationship between strain ($\varepsilon$), key parameters, and electron mobility is conceptually illustrated in Fig.~\ref{fig:ntype}a. This schematic outlines how lattice temperature ($T$), carrier concentration ($n$), impurity density ($n_{\mathrm{imp}}$), and dielectric environment influence three critical metrics: the unstrained electron mobility ($\mu_\mathrm{e0}$), the relative mobility enhancement ($\mu_\mathrm{e}/\mu_\mathrm{e0}$), and its strain derivative, the mobility enhancement rate ($\frac{\partial(\mu_\mathrm{e}/\mu_\mathrm{e0})}{\partial\varepsilon}$). We distinguish these quantities because they answer fundamentally different questions. The $\mu_\mathrm{e0}$ defines a material's baseline performance. Also, $\mu_\mathrm{e}/\mu_\mathrm{e0}$ reveals the intrinsic efficacy of strain by normalizing out material-specific scattering strengths. Finally, $\frac{\partial(\mu_\mathrm{e}/\mu_\mathrm{e0})}{\partial\varepsilon}$ provides a standardized sensitivity metric, essential for evaluating the robustness of strain tuning across diverse operational conditions.

Figure~\ref{fig:ntype}b presents a schematic representation of the conduction band structure evolution under biaxial strain, demonstrating the energy separation ($\Delta \mathrm{E}_{\mathrm{QK}}$) between the K and Q valleys. This strain-induced band modification represents a fundamental mechanism for controlling carrier transport in 2D semiconductors. Under tensile strain, the conduction band minimum (CBM) at the Q point shifts upward, while the K valley shifts downward; compressive strain reverses this trend, lowering the Q valley and raising the K valley. This phenomenon has been observed in both theoretical investigations \cite{kumar2024strain, junior2022first, wiktor2016absolute} and experimental studies \cite{kumar2024strain, yang2023strain}, reconfiguring the fundamental electronic landscape and predetermining the energy-dependent scattering phase space for carriers.

The quantitative evolution of the K--Q valley separation ($\Delta {\mathrm{E_{QK}}} = {\mathrm{E_Q}} - {\mathrm{E_K}}$) under biaxial strain is presented in Fig.~\ref{fig:ntype}c. At zero strain, the energy separations $\Delta \mathrm{E}_{\mathrm{QK}}$, are measured as 118 meV for MoS$_2$, 92 meV for MoSe$_2$, and 108 meV for WS$_2$, indicating that both valleys contribute significantly to electron transport in the unstrained condition, particularly at high carrier densities or for highly energetic electrons in the sample. Under applied strain, MoS$_2$, MoSe$_2$, and WS$_2$ exhibit pronounced and material-dependent valley shifts in their electronic band structures. Under tensile strain, the CBM at the Q point experiences an upward shift, while the K valley undergoes a concurrent downward shift. This counter-directional movement results in a net $\Delta \mathrm{E}_{\mathrm{QK}}$ of 234~meV/\%$\varepsilon$ for MoS$_2$, 211~meV/\%$\varepsilon$ for MoSe$_2$, and 278~meV/\%$\varepsilon$ for WS$_2$. In contrast, under compressive strain, the trend is reversed: the Q valley shifts downward while the K valley shifts upward, leading to a net $\Delta \mathrm{E}_{\mathrm{QK}}$ of 28~meV/\%$\varepsilon$, 43~meV/\%$\varepsilon$, and 39~meV/\%$\varepsilon$ for MoS$_2$, MoSe$_2$, and WS$_2$, respectively. These strain-induced modulations in valley energetics are in excellent agreement with prior first-principles calculations reported in the literatures~\cite{junior2022first, wiktor2016absolute}.

Figures~\ref{fig:ntype}d--i present the calculated density of states (DOS) across the K--Q pathway and intrinsic electron-phonon scattering rates for MoS$_2$, MoSe$_2$, and WS$_2$ under biaxial strain. The DOS in Figs.~\ref{fig:ntype}d--f reveals a systematic evolution with mechanical deformation. Compressive strain (blue) induces a pronounced enhancement of the DOS at lower energies, while tensile strain (red) suppresses it, particularly reducing the accessible states near the band edge. The unstrained condition (black) presents an intermediate profile. Concurrently, the intrinsic scattering rates (ADP, ODP, POP, IV, and PZ) in Figs.~\ref{fig:ntype}g--i show distinct material-dependent magnitudes and a clear strain response. Among the materials, WS$_2$ exhibits the lowest unstrained scattering rate. A key observation is the universal suppression of the total scattering rate under tensile strain and its enhancement under compressive strain across all three TMDs.

The interconnected trends observed in the DOS and scattering rates are fundamentally governed by the strain-mediated shifts in the conduction band valleys, as shown in Fig.~\ref{fig:ntype}c. The enhancement of the DOS under compressive strain results from the downward shift of the high effective-mass Q valley (see Supplementary Fig. 4b), which populates this valley with a high density of states. Conversely, the suppression of the DOS under tensile strain confirms the upward shift of the Q valley, energetically depopulating these states. This band modification controls the scattering landscape. The increased $\Delta \mathrm{E}_{\mathrm{QK}}$ under tensile strain raises the energy barrier for electrons to scatter from the K valley to the higher-energy Q valley. Given the finite phonon energies and thermal distributions at room temperature, this larger barrier exponentially reduces the probability of such IV transitions, leading to a notable decrease in the total scattering rate. The superior performance of WS$_2$ manifests in two key aspects. Its lowest baseline scattering rate originates from intrinsically weaker electron-phonon coupling, characterized by a low optical deformation potential ($D_{0}$) and high stretching modulus ($C_\mathrm{2D}$), as shown in Supplementary Tables 1 and 2. Simultaneously, its most pronounced response to strain is a direct consequence of its largest strain-induced $\Delta \mathrm{E}_{\mathrm{QK}}$ increase.

\begin{figure}[htp]
\centering
\includegraphics[width=0.95\textwidth]{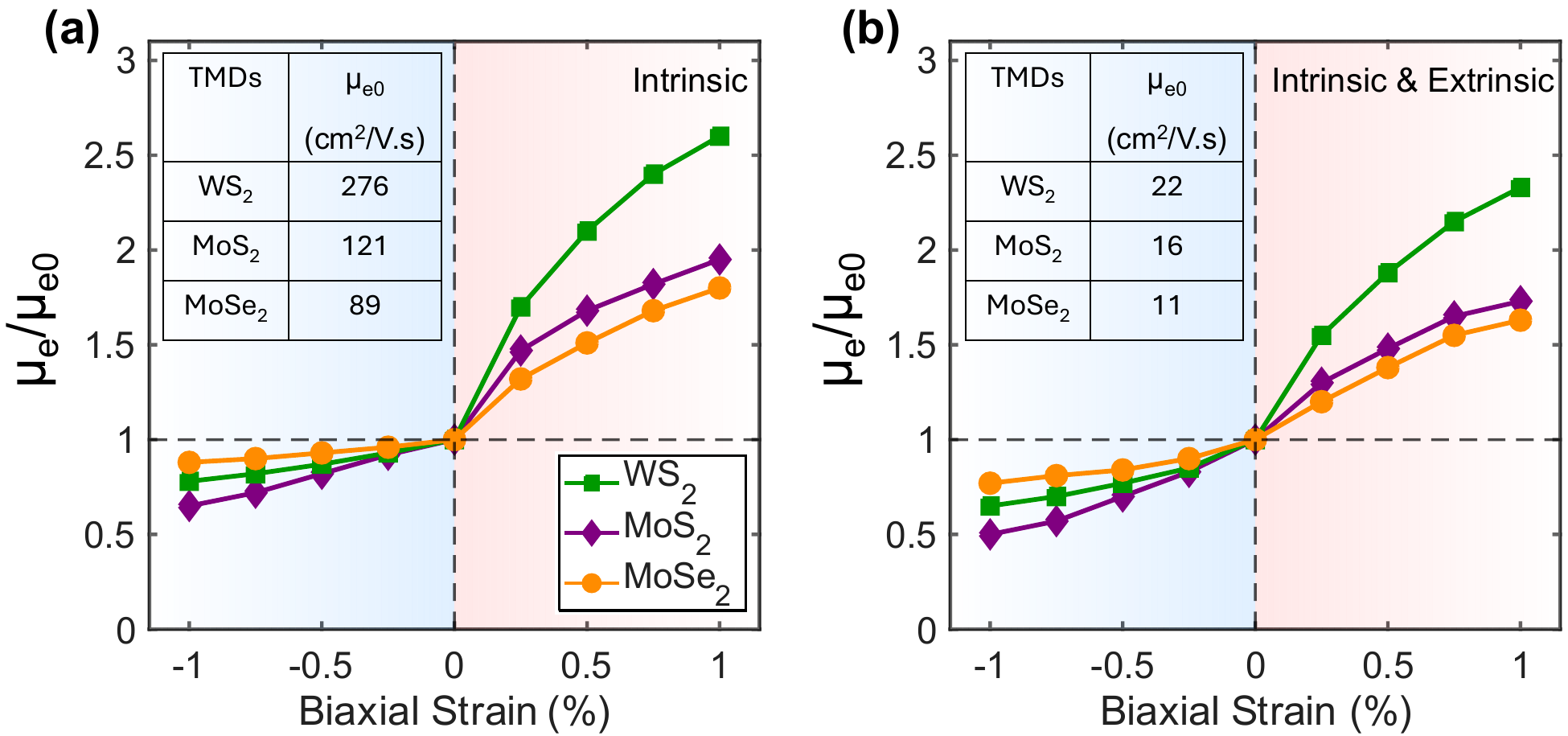}
\caption{\textbf{Enhanced electron mobility in n-type TMDs through strain engineering.} 
\textbf{a,} Intrinsic electron mobility enhancement considering ADP, ODP, POP, IV, and PZ scattering mechanisms. 
\textbf{b,} Total electron mobility enhancement incorporating both intrinsic and extrinsic effects (CI + SOP scattering). Results are shown for MoS$_2$ (purple), MoSe$_2$ (orange), and WS$_2$ (green) under biaxial strain at $T$ = 300 K, $n = 10^{13}$\,cm$^{-2}$, $n_{\text{imp}} = 5 \times 10^{12}$\,cm$^{-2}$, and SiO$_2$ dielectric environment. Tensile strain consistently enhances the electron mobility of all n-type TMDs, with WS$_2$ exhibiting the most significant improvement across both intrinsic and extrinsic scattering regimes.}
\label{fig:ntype_enhanc}
\end{figure}

To uncover the strain-induced modification of carrier transport from material-specific baseline properties, we frame our central results in terms of $\mu_\mathrm{e}/\mu_\mathrm{e0}$, rather than $\mu_\mathrm{e0}$. This representation directly quantifies the efficacy of strain engineering by normalizing out the varying scattering strengths between the TMDs considered, thereby allowing a clearer comparison of their strain response. Figure~\ref{fig:ntype_enhanc}a shows the intrinsic $\mu_\mathrm{e}/\mu_\mathrm{e0}$ as a function of biaxial strain, considering ADP, ODP, POP, IV, and PZ scattering mechanisms at $T$ = 300 K and $n = 10^{13}$ cm$^{-2}$. The baseline mobility, $\mu_\mathrm{e0}$, is estimated to be 121 cm$^2$/V$\cdot$s  for MoS$_2$, 89 cm$^2$/V$\cdot$s for MoSe$_2$, and 276 cm$^2$/V$\cdot$s for WS$_2$.  We computed the carrier mobility using equation (\ref{eq:final_mu_main}) that integrates the calculated scattering rates with group velocities and densities of states obtained from our first-principles calculations. The significantly higher mobility in WS$_2$ directly correlates with its lower intrinsic scattering rate observed in Fig.~\ref{fig:ntype}i. Under tensile strain conditions, the intrinsic $\mu_\mathrm{e}/\mu_\mathrm{e0}$ are 1.96/\%$\varepsilon$ for MoS$_2$, 1.82/\%$\varepsilon$ for MoSe$_2$, and 2.63/\%$\varepsilon$ for WS$_2$. The superior mobility enhancement in WS$_2$ results from its largest increase in $\Delta \mathrm{E}_{\mathrm{QK}}$ (170 meV/\%strain), which most effectively suppresses IV scattering.

Figure~\ref{fig:ntype_enhanc}b presents the $\mu_\mathrm{e}/\mu_\mathrm{e0}$ when including extrinsic effects, specifically CI scattering and SO phonon interactions at $T$ = 300 K, $n = 10^{13}$\,cm$^{-2}$, $n_{\text{imp}} = 5 \times 10^{12}$\,cm$^{-2}$, and SiO$_2$ dielectric environment. The incorporation of these extrinsic scattering mechanisms notably reduces the $\mu_\mathrm{e0}$ to 16 cm$^2$/V$\cdot$s for MoS$_2$, 11 cm$^2$/V$\cdot$s for MoSe$_2$, and 22 cm$^2$/V$\cdot$s for WS$_2$. This reduction occurs because charged impurities, typically located near the interface between the TMD monolayer and the substrate, create long-range Coulomb potentials that deflect carriers and disrupt their transport. Additionally, SO phonons originating from polar substrates interact with carriers through remote coupling, introducing inelastic scattering that becomes particularly significant at room temperature. These extrinsic scattering mechanisms, which are absent in intrinsic calculations, substantially degrade carrier mobility and often dominate transport in realistic device setups. The $\mu_\mathrm{e}/\mu_\mathrm{e0}$ in the presence of extrinsic effects reduce to 1.71/\%$\varepsilon$ for MoS$_2$, 1.62/\%$\varepsilon$ for MoSe$_2$, and 2.31/\%$\varepsilon$ for WS$_2$. Crucially, despite substantially degrading absolute mobility, these extrinsic mechanisms preserve the relative strain enhancement factor, as they predominantly scatter low-energy carriers while strain uniformly shifts the entire band structure, maintaining the proportional improvement. This reduction occurs because CI and SOP scattering mechanisms are less sensitive to strain-induced changes in the electronic band structure. CI scattering arises from long-range Coulomb potentials that, while band-structure dependent, primarily affect low-energy electrons due to their electrostatic nature, with diminished impact on higher-energy states. SOP scattering mainly depends on substrate properties, but similarly couples more strongly with carriers near the band edge. Despite this reduction, WS$_2$ maintains the highest enhancement due to its intrinsically weaker electron-phonon coupling and reduced sensitivity to substrate-induced phonon modes.

Our theoretical framework not only explains the enhancement mechanisms within the experimentally accessible strain range but also predicts the robustness of this phenomenon at significantly larger strain levels. Our predictions reveal that $\mu_\mathrm{e}/\mu_\mathrm{e0}$ persists under large strain (up to 5\%), as shown in Supplementary Fig. 7. However, the $\mu_\mathrm{e}/\mu_\mathrm{e0}$ diminishes at higher strain levels compared to the lower strains. This reduction correlates with a decreased rate of change in $\Delta \mathrm{E}_{\mathrm{QK}}$ at higher deformations; the conduction band valleys shift less notably per unit strain, resulting in a more gradual suppression of IV scattering. This trend is consistent for both intrinsic (Supplementary Fig. 7a) and extrinsic cases (Supplementary Fig. 7b), confirming that while strain remains beneficial, its effectiveness becomes less pronounced in the high-strain regime.

\begin{figure}[htp]
\centering
\includegraphics[width=0.97\textwidth]{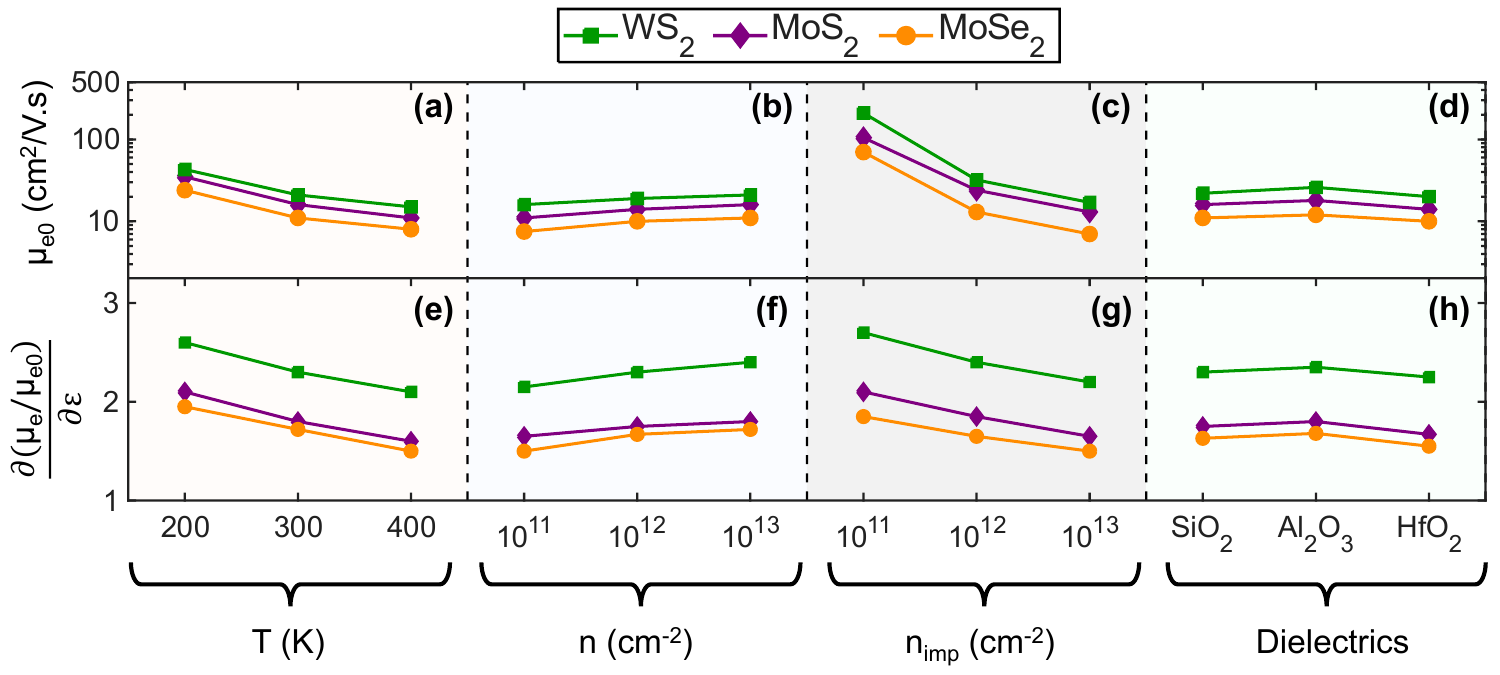}
\caption{\textbf{Parametric analysis of intrinsic and extrinsic factors influencing strain-induced electron mobility enhancement in n-type TMDs.} 
Unstrained initial electron mobility versus \textbf{a,} temperature, \textbf{b,} carrier concentration, \textbf{c,} impurity concentration, and \textbf{d,} dielectric environment. Mobility enhancement factor per percent strain (evaluated at 1\% biaxial tensile strain) versus \textbf{e,} temperature, \textbf{f,} carrier concentration, \textbf{g,} impurity concentration, and \textbf{h,} dielectric environment.  Unless otherwise specified, analyses are performed at 300~K, with $n = 10^{13}$\,cm$^{-2}$, $n_{\text{imp}} = 5 \times 10^{12}$\,cm$^{-2}$, and a SiO$_2$ dielectric. Results are shown for MoS$_2$ (purple), MoSe$_2$ (orange), and WS$_2$ (green). The tensile strain-induced electron mobility enhancement trend remains robust across orders-of-magnitude variation in all parameters, with WS$_2$ consistently showing the best performance.}
\label{fig:ntype_param}
\end{figure}

Having established the significant enhancement potential under ideal conditions, we next evaluate its robustness in Fig.~\ref{fig:ntype_param} for device applications by quantifying the strain sensitivity, $\frac{\partial(\mu_\mathrm{e}/\mu_\mathrm{e0})}{\partial\varepsilon}$. Defined as the slope of the enhancement curves in Fig.~\ref{fig:ntype_enhanc}b, this parameter provides a standardized metric to compare the impact of various practical parameters on the effectiveness of strain tuning, independent of their individual effects on the baseline mobility $\mu_\mathrm{e0}$. Figures~\ref{fig:ntype_param}a-d present the variation of $\mu_\mathrm{e0}$ with temperature ($T$ = 200--400 K), carrier density ($n$ = 10$^{11}$--10$^{13}$\,cm$^{-2}$), impurity density ($n_{\text{imp}}$ = 10$^{11}$--10$^{13}$\,cm$^{-2}$), and dielectric environment (SiO$_2$, Al$_2$O$_3$, HfO$_2$). Additionally, Figs.~\ref{fig:ntype_param}e-h evaluate the sensitivity, $\frac{\partial(\mu_\mathrm{e}/\mu_\mathrm{e0})}{\partial\varepsilon}$ to these same parameters at 1\% biaxial tensile strain.

Temperature suppresses both the baseline mobility and its strain enhancement. The monotonic decrease in $\mu_\mathrm{e0}$ with rising temperature (see Fig.~\ref{fig:ntype_param}a) stems from intensified electron-phonon scattering. Elevated thermal energy increases the population of acoustic and optical phonons, while a broadened Fermi-Dirac distribution promotes the occupation of higher-energy side valleys, enhancing IV scattering. This thermal agitation also minimizes $\frac{\partial(\mu_\mathrm{e}/\mu_\mathrm{e0})}{\partial\varepsilon}$ (see Fig.~\ref{fig:ntype_param}e). Although tensile strain increases the $\Delta \mathrm{E}_{\mathrm{QK}}$, the broader electron distribution at high $T$ allows more carriers to populate the Q valley, thereby weakening the impact of strain on suppressing IV scattering.

Carrier density exhibits a synergistic influence, enhancing both mobility and its strain response. Figure~\ref{fig:ntype_param}b shows $\mu_\mathrm{e0}$ increasing with carrier concentration, a direct consequence of enhanced electrostatic screening of charged impurity potentials. Although this trend may saturate at very high densities, increased carrier screening remains the dominant mechanism within the studied range. This screening effect also clarifies the intrinsic scattering landscape, which is more susceptible to strain modulation. Consequently, $\frac{\partial(\mu_\mathrm{e}/\mu_\mathrm{e0})}{\partial\varepsilon}$ rises with carrier density (see Fig.~\ref{fig:ntype_param}f), as the suppression of extrinsic impurity scattering allows strain-engineered suppression of IV scattering to dominate the transport behavior.

The influence of impurity density has a predictably adverse effect on performance. A higher $n_{\text{imp}}$ introduces more scattering centers, and the resulting long-range Coulomb disorder notably reduces $\mu_\mathrm{e0}$ (see Fig.~\ref{fig:ntype_param}c). When charged impurity scattering becomes the dominant mobility-limiting mechanism, the effect of strain is marginalized. This leads to a pronounced reduction in $\frac{\partial(\mu_\mathrm{e}/\mu_\mathrm{e0})}{\partial\varepsilon}$ (see Fig.~\ref{fig:ntype_param}g), as strain-induced modulations of intrinsic phonon processes contribute less to the total mobility.

Finally, the role of the dielectric environment reveals a critical insight for device engineering. The substrate influences $\mu_\mathrm{e0}$ (see Fig.~\ref{fig:ntype_param}d) via SOP scattering, a process governed by a nuanced balance of phonon energy, occupation, and coupling strength. Dielectrics like SiO$_2$ and HfO$_2$ exhibit significant scattering due to their lower phonon energies, higher phonon occupancy, and stronger interfacial coupling. In contrast, Al$_2$O$_3$ shows slightly reduced scattering and consequently offers better mobility compared to the other two dielectrics, benefiting from its comparatively weaker interfacial coupling strength. As shown in Fig.~\ref{fig:ntype_param}h, the $\frac{\partial(\mu_\mathrm{e}/\mu_\mathrm{e0})}{\partial\varepsilon}$ exhibits a weak dependence on the dielectric environment. The data indicate that SOP effects have minimal impact on $\frac{\partial(\mu_\mathrm{e}/\mu_\mathrm{e0})}{\partial\varepsilon}$. This suggests that band structure modifications under strain dominate over dielectric-dependent scattering mechanisms.

Throughout these parametric variations, WS$_2$ consistently demonstrates the highest $\frac{\partial(\mu_\mathrm{e}/\mu_\mathrm{e0})}{\partial\varepsilon}$ under tensile strain. This superior performance stems from its favorable combination of large $\Delta \mathrm{E}_{\mathrm{QK}}$ increase, high $C_\mathrm{2D}$, low $D_{0}$, and reduced sensitivity to extrinsic scattering mechanisms. Even in the presence of extrinsic effects, such as CI and SOP scattering, WS$_2$ maintains an advantage due to weaker electron-phonon interaction and lower sensitivity to substrate-induced phonon modes. These intrinsic and extrinsic factors together make WS$_2$ the most responsive to strain engineering, resulting in enhanced carrier transport and the highest overall mobility gains among the TMDs considered. 
Furthermore, the consistent enhancement of electron mobility through strain in all three n-type TMDs (MoS$_2$,  MoSe$_2$, and WS$_2$) across varying temperatures, carrier densities, impurity concentrations, and dielectric environments underscores strain engineering as a robust and broadly applicable strategy for improving electron transport in devices based on 2D n-type TMDs.

\subsection*{Hole mobility enhancement}

\begin{figure}[htp]
\centering
\includegraphics[width=0.97\textwidth]{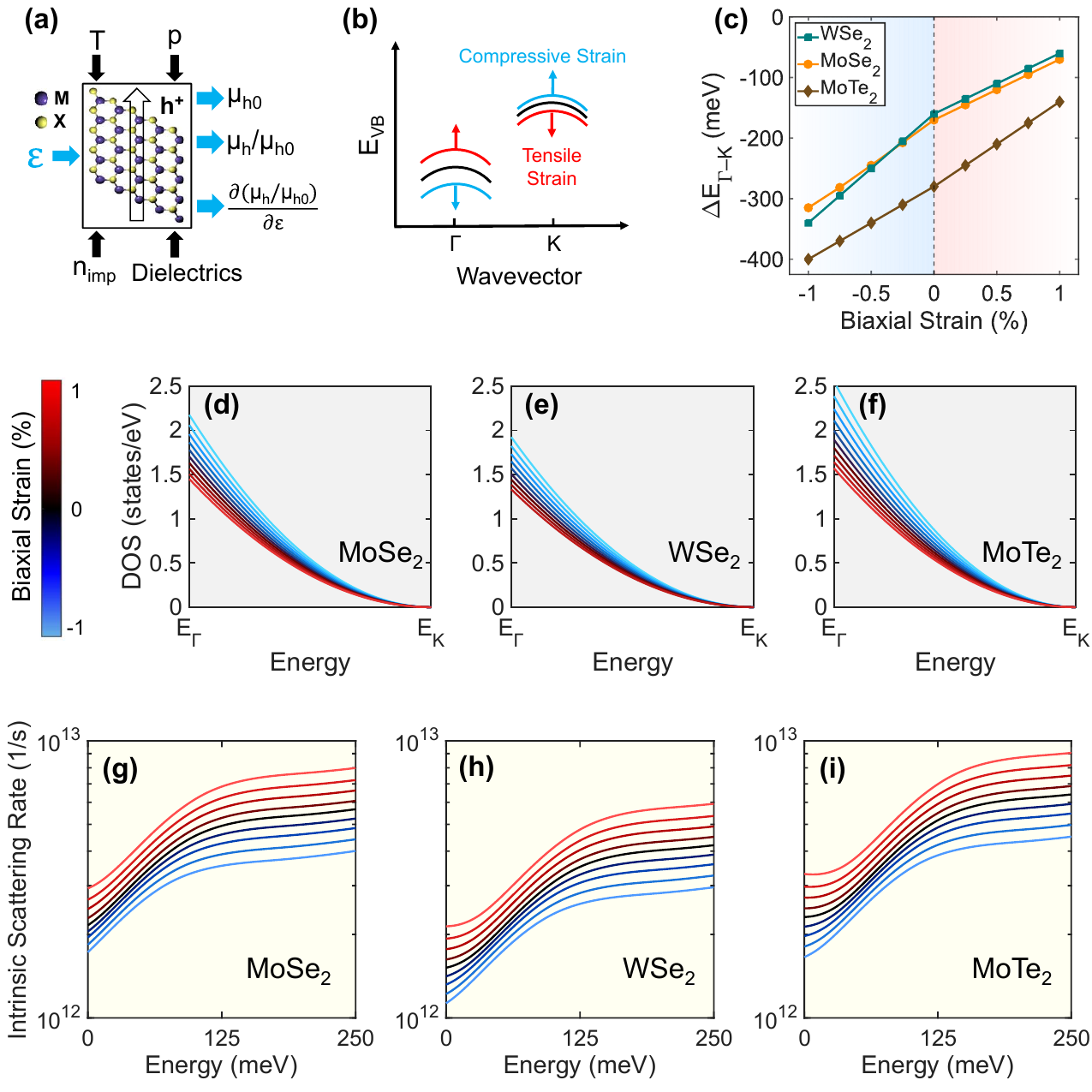}
\caption{\textbf{Mechanisms of strain-induced hole mobility enhancement in p-type TMDs.} 
\textbf{a,} Schematic overview of the key parameters governing hole mobility in p-type TMDs, highlighting the functional relationship between mobility enhancement and applied strain.  
\textbf{b,} Schematic of the valence band structure showing energy separation between $\Gamma$ and K valleys ($\Delta \mathrm{E}_{\mathrm{\Gamma K}}$) under biaxial strain. \textbf{c,} Evolution of the energy separation $\Delta \mathrm{E}_{\mathrm{\Gamma K}}$ under biaxial strain for MoSe$_2$ (orange), WSe$_2$ (teal), and MoTe$_2$ (brown). First-principles calculated density of states versus valence band energy along the $\Gamma$--K direction under varying strain for \textbf{d,} MoSe$_2$, \textbf{e,} WSe$_2$, and \textbf{f,} MoTe$_2$. Computed intrinsic scattering rates versus carrier energy under varying strain for \textbf{g,} MoSe$_2$, \textbf{h,} WSe$_2$, and \textbf{i,} MoTe$_2$, showing suppression of scattering under compressive strain.}
\label{fig:ptype}
\end{figure}

Here we present a comprehensive analysis of hole mobility enhancement in monolayer p-type TMDs--MoSe$_2$, WSe$_2$, and MoTe$_2$, under biaxial strain. The conceptual framework linking key model parameters to hole mobility is depicted in Fig.~\ref{fig:ptype}a. This schematic outlines the impact of key factors, $T$, $p$ (hole concentration), $n_{\mathrm{imp}}$, dielectric environment, and applied $\varepsilon$, on the hole mobility. The core relationship captures the variation of the unstrained hole mobility ($\mu_\mathrm{h0}$), the relative mobility enhancement ($\mu_\mathrm{h}/\mu_\mathrm{h0}$), and the strain-induced enhancement rate $\left(\frac{\partial(\mu_\mathrm{h}/\mu_\mathrm{h0})}{\partial\varepsilon}\right)$ as functions of strain.

Figure~\ref{fig:ptype}b illustrates the valence band evolution under biaxial strain, showing the energy offset ($\Delta \mathrm{E}_{\mathrm{\Gamma K}}$) between the $\Gamma$ and K valleys. Under compressive strain, the $\Gamma$ valley undergoes a notable downward shift, while the K valley shifts upward minimally, resulting in an increased IV transition barrier. Under tensile strain, the trend reverses, with the $\Gamma$ valley shifting upward and reducing $\Delta \mathrm{E}_{\mathrm{\Gamma K}}$. This band modification mechanism, observed in both theoretical investigations \cite{wiktor2016absolute, cheng2020using} and experimental study \cite{shen2016strain}.

The quantitative evolution of the $\Gamma$--K valley separation ($\Delta {\mathrm{E_{\Gamma K}}} = {\mathrm{E_\Gamma}} - {\mathrm{E_K}}$) under biaxial strain is presented in Fig.~\ref{fig:ptype}c. At zero strain, the energy separations $\Delta \mathrm{E}_{\mathrm{\Gamma K}}$ measure 166 meV for MoSe$_2$, 157 meV for WSe$_2$, and 282 meV for MoTe$_2$, 
indicating that both the $\Gamma$ and K valleys can contribute to hole transport, especially at elevated carrier densities or for high-energy holes in the sample. Under compressive strain, the $\Gamma$ valley shifts downward while the K valley shifts upward only marginally, resulting in net $\Delta \mathrm{E}_{\mathrm{\Gamma K}}$ of 315 meV/\%$\varepsilon$ for MoSe$_2$, 341 meV/\%$\varepsilon$ for WSe$_2$, and 398 meV/\%$\varepsilon$ for MoTe$_2$. In contrast, tensile strain reverses this trend, raising the $\Gamma$ valley and lowering the K valley. These strain-induced modulations in valley energetics show excellent agreement with previous first-principles calculations reported in the literatures~\cite {cheng2020using, wiktor2016absolute}.

Figures~\ref{fig:ptype}d–f present the calculated DOS along the $\Gamma$--K valence pathway, revealing the strain-induced tuning of the valence band valleys. The suppression of the DOS under compressive strain signifies that the heavy-mass $\Gamma$ valley (Supplementary Fig. 4d) shifts downward, away from the VBM. This energetically isolates the $\Gamma$ valley and funnels the entire hole population into the lightweight K valley at the VBM. In contrast, tensile strain enhances the DOS by raising the $\Gamma$ valley closer to the K point, populating these heavy-mass states. This mechanism directly explains the suppressed scattering observed under compressive strain in Figs.~\ref{fig:ptype}g–i. Among the three p-type TMDs, WSe$_2$ exhibits the lowest unstrained intrinsic scattering rate due to its high ${C_\mathrm{2D}}$ (Supplementary Table 1) and low ${D_{0}}$ (Supplementary Table 2). These intrinsic properties make WSe$_2$ particularly resistant to hole-phonon scattering. Despite possessing the largest $\Delta \mathrm{E}_{\mathrm{\Gamma K}}$ (282 meV) in its unstrained state, MoTe$_2$ exhibits the lowest hole mobility. This results from its characteristically lower ${C_\mathrm{2D}}$ (Supplementary Table 1) and stronger ${D_{0}}$ (Supplementary Table 2), which dominate its carrier transport properties before the application of strain.

A detailed understanding of the strain dependence of hole scattering rates is crucial for effectively controlling transport in the material. With increasing compressive strain, the scattering rate decreases significantly across all materials due to suppressed IV scattering. The increased $\Delta \mathrm{E}_{\mathrm{\Gamma K}}$ requires more energy for holes to scatter from the $\Gamma$ valley to the higher-energy K valley. Given finite phonon energies and thermal distributions at room temperature, this larger energy barrier reduces the transition probability. Conversely, tensile strain reduces $\Delta \mathrm{E}_{\mathrm{\Gamma K}}$, enhancing IV scatterings in the material. This relationship between valley separation and scattering probability explains the observed trends in both Figs.~\ref{fig:ptype}c and~\ref{fig:ptype}g-i.

\begin{figure}[htp]
\centering
\includegraphics[width=0.95\textwidth]{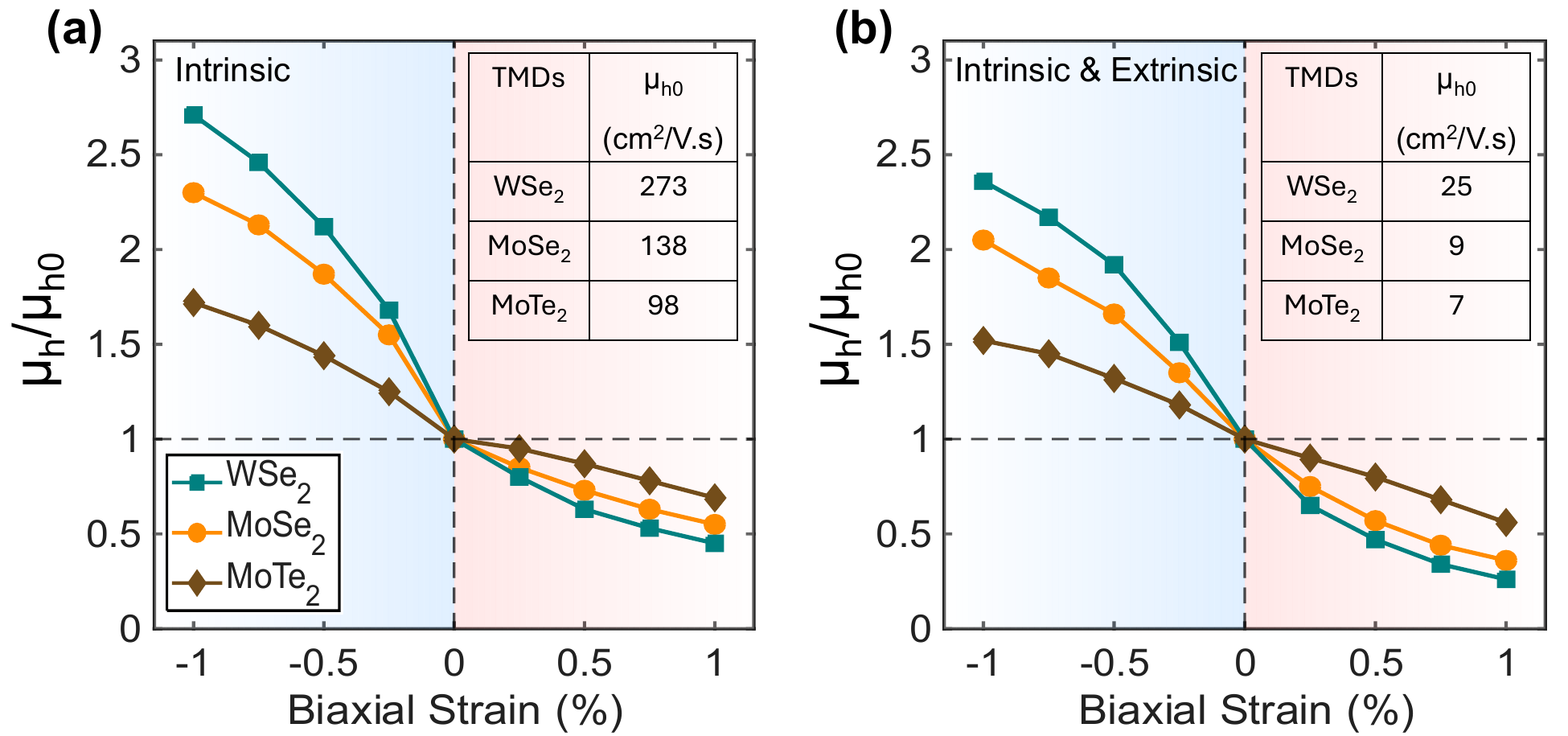}
\caption{\textbf{Enhanced hole mobility in p-type TMDs through strain engineering.} 
\textbf{a,} Intrinsic hole mobility enhancement considering ADP, ODP, POP, IV, and PZ scattering. 
\textbf{b,} Total hole mobility enhancement incorporating both intrinsic and extrinsic effects (CI + SOP scattering). Results are shown for MoSe$_2$ (orange), WSe$_2$ (teal), and MoTe$_2$ (brown) under biaxial strain at $T$ = 300 K, $p = 10^{13}$\,cm$^{-2}$, $n_{\text{imp}} = 5 \times 10^{12}$\,cm$^{-2}$, and SiO$_2$ dielectric environment. Compressive strain consistently enhances the hole mobility of all p-type TMDs, with WSe$_2$ exhibiting the most significant improvement across both intrinsic and extrinsic scattering regimes.}
\label{fig:ptype_enhanc}
\end{figure}

Figure~\ref{fig:ptype_enhanc}a shows the intrinsic hole mobility enhancement with strain, considering ADP, ODP, POP, IV, and PZ scattering at $T$ = 300 K and $p = 10^{13}$\,cm$^{-2}$. The $\mu_\mathrm{h0}$ calculate as 138 cm$^2$/V$\cdot$s for MoSe$_2$, 273 cm$^2$/V$\cdot$s for WSe$_2$, and 98 cm$^2$/V$\cdot$s for MoTe$_2$. The higher mobility in WSe$_2$ directly correlates with its lower intrinsic scattering rate observed in Fig.~\ref{fig:ptype}h. Under compressive strain, the $\mu_\mathrm{h}/\mu_\mathrm{h0}$ are 2.34/\%$\varepsilon$ for MoSe$_2$, 2.71/\%$\varepsilon$ for WSe$_2$, and 1.68/\%$\varepsilon$ for MoTe$_2$. Although MoTe$_2$ has the largest $\Delta \mathrm{E}_{\mathrm{\Gamma K}}$ under no strain, its smaller compressive strain-induced increase in $\Delta \mathrm{E}_{\mathrm{\Gamma K}}$ (116 meV/\%$\varepsilon$) and stronger hole-phonon coupling limit its enhancement potential. The superior enhancement in WSe$_2$ results from its largest increase in $\Delta \mathrm{E}_{\mathrm{\Gamma K}}$ (184 meV/\%$\varepsilon$), which most effectively suppresses IV scattering under compressive strain.

Figure~\ref{fig:ptype_enhanc}b includes extrinsic effects from CI scattering and SOP interactions at $T$ = 300 K, $p = 10^{13}$\,cm$^{-2}$, $n_{\text{imp}} = 5 \times 10^{12}$\,cm$^{-2}$, and SiO$_2$ dielectric environment. These mechanisms reduce $\mu_\mathrm{h0}$ to 9 cm$^2$/V$\cdot$s for MoSe$_2$, 25 cm$^2$/V$\cdot$s for WSe$_2$, and 7 cm$^2$/V$\cdot$s for MoTe$_2$. Charged impurities near the TMD-substrate interface create long-range Coulomb potentials that deflect holes, while SOP from polar substrates introduces inelastic scattering through remote coupling. The $\mu_\mathrm{h}/\mu_\mathrm{h0}$ reduces to 2.14/\%$\varepsilon$ for MoSe$_2$, 2.37/\%$\varepsilon$ for WSe$_2$, and 1.52/\%$\varepsilon$ for MoTe$_2$ because CI and SOP scattering are less sensitive to strain-induced changes to the material's electronic structure. Despite this reduction, WSe$_2$ maintains the highest enhancement due to its weaker hole-phonon coupling and reduced sensitivity to substrate-induced phonon modes.

Our computational framework predicts that $\mu_\mathrm{h}/\mu_\mathrm{h0}$ persists under large strain (up to 5\%), as shown in Supplementary Fig. 8. However, the $\mu_\mathrm{h}/\mu_\mathrm{h0}$ diminishes at higher strain levels due to reduced rates of $\Delta \mathrm{E}_{\mathrm{\Gamma K}}$ change; the valence band valleys shift less notably per unit strain, resulting in more gradual suppression of IV scattering. This trend holds for both intrinsic (Supplementary Fig. 8a) and extrinsic cases (Supplementary Fig. 8b), confirming that while strain remains beneficial, its effectiveness decreases in the high-strain regime.

\begin{figure}[htp]
\centering
\includegraphics[width=0.97\textwidth]{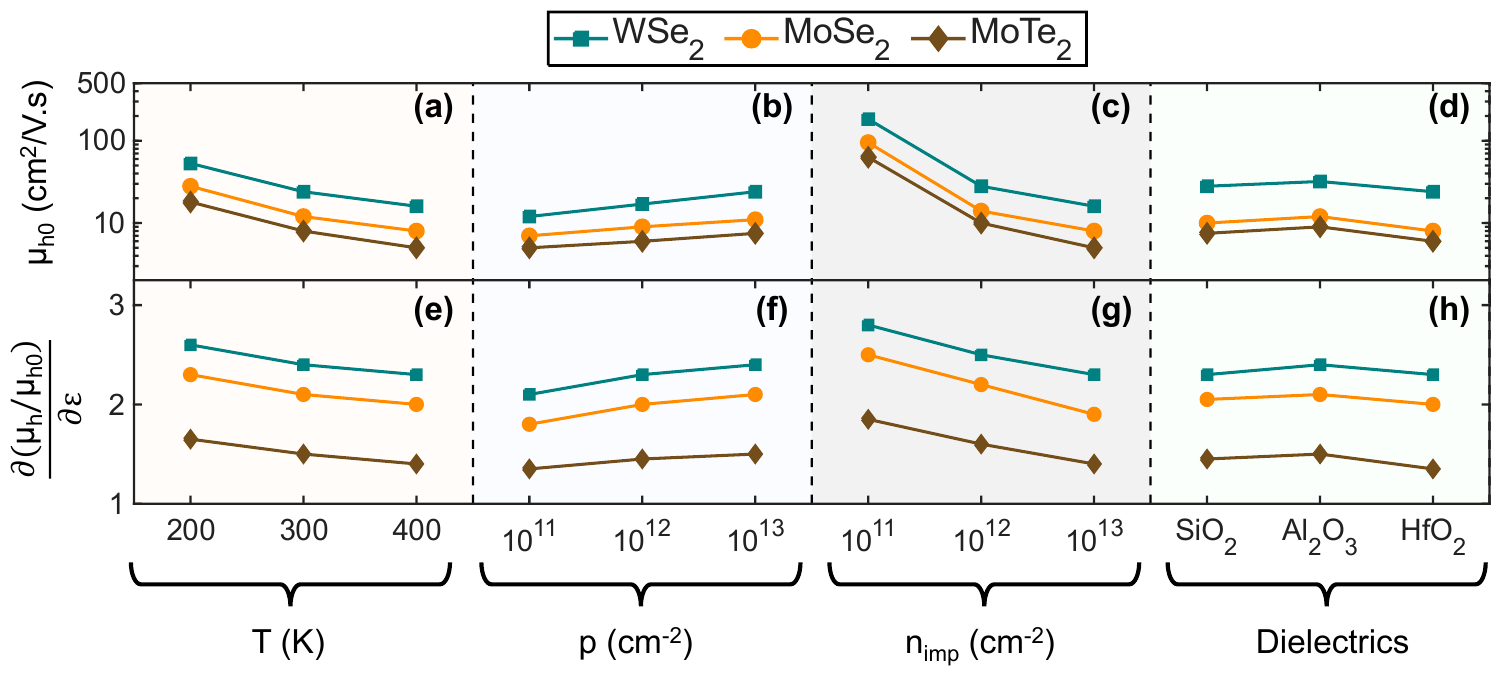}
\caption{\textbf{Parametric analysis of intrinsic and extrinsic factors influencing strain-induced hole mobility enhancement in p-type TMDs.} 
Unstrained initial hole mobility versus \textbf{a,} temperature, \textbf{b,} carrier concentration, \textbf{c,} impurity concentration, and \textbf{d,} dielectric environment. Mobility enhancement factor per percent strain (evaluated at –1\% biaxial compressive strain) versus \textbf{e,} temperature, \textbf{f,} carrier concentration, \textbf{g,} impurity concentration, and \textbf{h,} dielectric environment. Unless otherwise specified, analyses are performed at 300~K, with $p = 10^{13}$\,cm$^{-2}$, $n_{\text{imp}} = 5 \times 10^{12}$\,cm$^{-2}$, and a SiO$_2$ dielectric. Results are shown for MoSe$_2$ (orange), WSe$_2$ (teal), and MoTe$_2$ (brown). The compressive strain-induced hole mobility enhancement trend remains robust across orders-of-magnitude variation in all parameters, with WSe$_2$ consistently showing the best performance.}
\label{fig:ptype_param}
\end{figure}

We further analyzed the robustness of strain-induced hole mobility enhancement across various parameters and extrinsic factors (see Fig.~\ref{fig:ptype_param}). The $\mu_\mathrm{h0}$ and  $\frac{\partial(\mu_\mathrm{h}/\mu_\mathrm{h0})}{\partial\varepsilon}$ exhibit systematic and interpretable trends. Figures~\ref{fig:ptype_param}a-d show $\mu_\mathrm{h0}$ variations with temperature ($T$ = 200--400 K), carrier density ($p$ = 10$^{11}$--10$^{13}$\,cm$^{-2}$), impurity density ($n_{\text{imp}}$ = 10$^{11}$--10$^{13}$\,cm$^{-2}$), and dielectric environment (SiO$_2$, Al$_2$O$_3$, HfO$_2$). Furthermore, Figs.~\ref{fig:ptype_param}e-h show the $\frac{\partial(\mu_\mathrm{h}/\mu_\mathrm{h0})}{\partial\varepsilon}$ under the same conditions (evaluated at -1\% biaxial compressive strain).

The  $\mu_\mathrm{h0}$ decreases with rising temperature (see Fig.~\ref{fig:ptype_param}a) due to enhanced hole-phonon scattering from increased phonon populations, and correspondingly, the strain sensitivity $\frac{\partial(\mu_\mathrm{h}/\mu_\mathrm{h0})}{\partial\varepsilon}$ diminishes (see Fig.~\ref{fig:ptype_param}e) as thermal broadening allows more holes to access the K valley despite the increased $\Delta \mathrm{E}_{\mathrm{\Gamma K}}$, thereby weakening the strain effect. Conversely, $\mu_\mathrm{h0}$ increases with carrier density (see Fig.~\ref{fig:ptype_param}b) owing to improved screening of Coulomb impurities; this suppression of extrinsic scattering allows strain-modulated intrinsic processes to dominate, leading to a concurrent increase in $\frac{\partial(\mu_\mathrm{h}/\mu_\mathrm{h0})}{\partial\varepsilon}$ (see Fig.~\ref{fig:ptype_param}f). As expected, a higher impurity density reduces $\mu_\mathrm{h0}$ (see Fig.~\ref{fig:ptype_param}c) by introducing more scattering centers, which also flattens the enhancement trend, causing $\frac{\partial(\mu_\mathrm{h}/\mu_\mathrm{h0})}{\partial\varepsilon}$ to decrease (see Fig.~\ref{fig:ptype_param}g) as charged impurity scattering becomes dominant. Finally, dielectrics with low $\hbar\omega_{\mathrm{SO}}$ (e.g., HfO$_2$) cause stronger scattering (low $\mu_\mathrm{h0}$) than those with high $\hbar\omega_{\mathrm{SO}}$ modes (e.g., Al$_2$O$_3$) due to higher phonon occupation at room temperature (see Fig.~\ref{fig:ptype_param}d). The $\frac{\partial(\mu_\mathrm{h}/\mu_\mathrm{h0})}{\partial\varepsilon}$ shows minimal dependence on the dielectric environment and associated SOP effects (see Fig.~\ref{fig:ptype_param}h).

Throughout these variations, WSe$_2$ consistently shows the highest enhancement, and the compressive strain effect remains robust. The notable performance of WSe$_2$ stems from its optimal combination of large $\Delta \mathrm{E}_{\mathrm{\Gamma K}}$ increase, high ${C_\mathrm{2D}}$, low ${D_{0}}$, and reduced sensitivity to extrinsic scattering mechanisms. Even in the presence of extrinsic scattering sources, WSe$_2$ maintains an advantage due to its weaker hole-phonon interaction and lower sensitivity to SOP. These factors make WSe$_2$ the most responsive to strain engineering, yielding the highest hole mobility gains among p-type TMDs. The robustness of strain-dependent hole mobility enhancement across temperature, carrier density, impurity levels, and dielectric environments establishes strain engineering as a universal approach for enhancing the performance of TMD-based devices that rely on hole transport.

\section*{Discussion}

\begin{figure}[htp]
\centering
\includegraphics[width=0.97\textwidth]{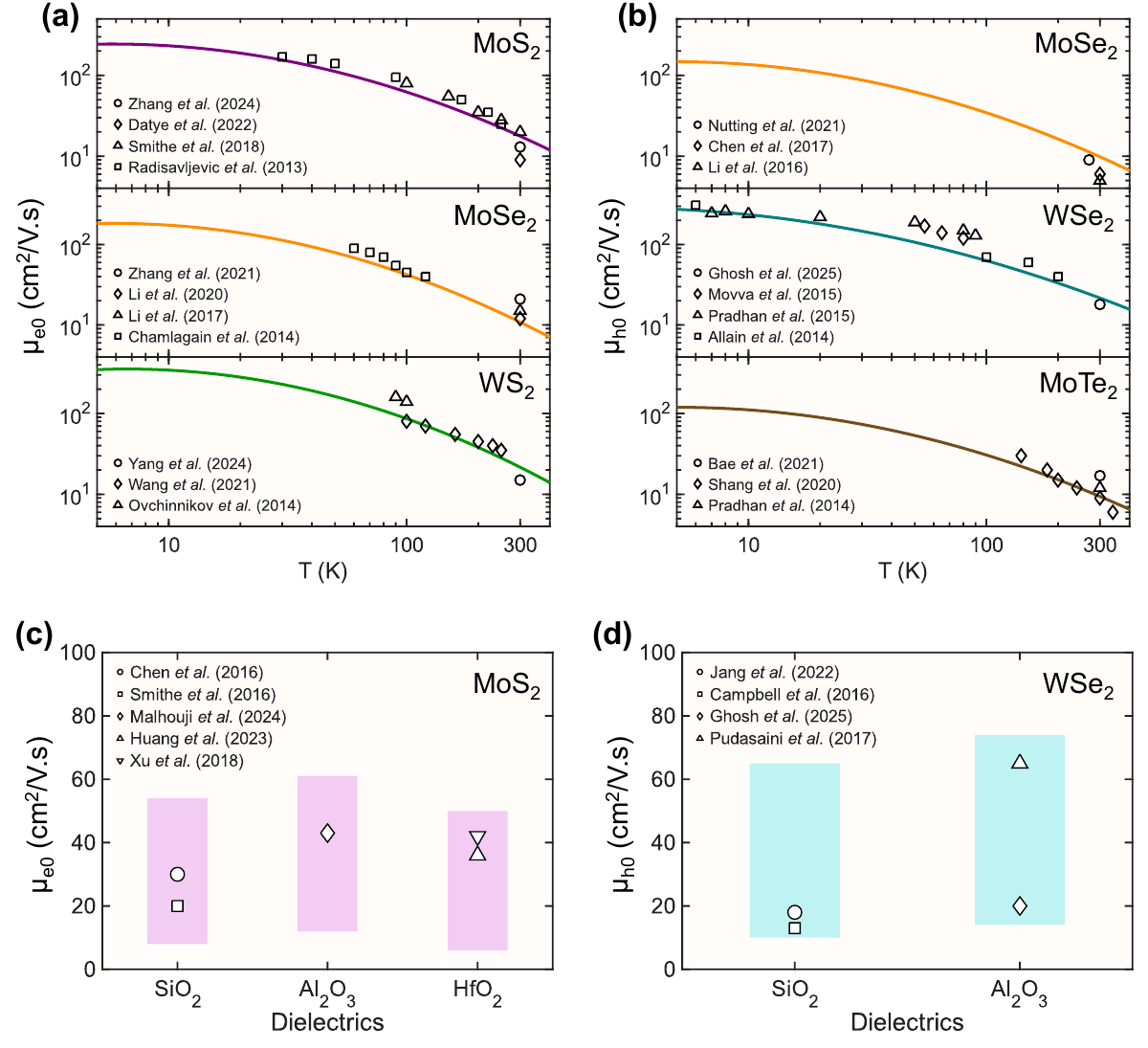}
\caption{\textbf{Comprehensive validation of unstrained carrier mobility in 2D TMDs.} 
\textbf{a,} Calculated unstrained electron mobility versus temperature for monolayer n-type TMDs, incorporating intrinsic and extrinsic scattering mechanisms ($n = 10^{13}\,\mathrm{cm}^{-2}$, $n_{\text{imp}} = 5\times10^{12}\,\mathrm{cm}^{-2}$, SiO$_2$ dielectric), compared with experimental data: MoS$_2$ \cite{zhang2024enhancing, datye2022strain, smithe2018high, radisavljevic2013mobility}, MoSe$_2$ \cite{zhang2021rapid, li2020sub, li2017scalable, chamlagain2014mobility}, and WS$_2$ \cite{yang2024biaxial, wang2021electron, ovchinnikov2014electrical}. 
\textbf{b,} Calculated unstrained hole mobility versus temperature for monolayer p-type TMDs under identical simulation conditions ($p = 10^{13}\,\mathrm{cm}^{-2}$), validated against experiments: MoSe$_2$ \cite{nutting2021electrical, chen2017highly, li2016isoelectronic}, WSe$_2$ \cite{ghosh2025high, movva2015high, pradhan2015hall, allain2014electron}, and MoTe$_2$ \cite{bae2021mote2, shang2020situ, pradhan2014field}. 
\textbf{c,} Dielectric-dependent unstrained electron mobility for MoS$_2$ at room temperature ($n = 10^{11}\text{--}10^{13}\,\mathrm{cm}^{-2}$, $n_{\text{imp}} = 10^{12}\text{--}10^{13}\,\mathrm{cm}^{-2}$), showing our calculated range (purple) across SiO$_2$, Al$_2$O$_3$, and HfO$_2$ dielectrics encompasses experimental data from \cite{chen2016chemical, smithe2016intrinsic, mahlouji2024influence, huang2023performance, xu2018effects}.
\textbf{d,} Dielectric-dependent unstrained hole mobility for WSe$_2$ at room temperature ($p = 10^{11}\text{--}10^{13}\,\mathrm{cm}^{-2}$, $n_{\text{imp}} = 10^{12}\text{--}10^{13}\,\mathrm{cm}^{-2}$), with our calculated range (teal) for SiO$_2$ and Al$_2$O$_3$ dielectrics showing excellent agreement with experiments from \cite{jang2022fermi, campbell2016field, pudasaini2017high, ghosh2025high}.}
\label{fig:exp_validation}
\end{figure}

Our investigation establishes biaxial strain engineering as a powerful and universal strategy for enhancing carrier mobility in both n- and p-type monolayer TMDs. The predictive power of our multi-scale computational framework is demonstrated through comprehensive validation across multiple fronts. First, our calculations of $\mu_\mathrm{e0}$ and $\mu_\mathrm{h0}$ show excellent agreement with a wide range of experimental measurements at varying temperatures. For n-type TMDs, our computed mobility for MoS$_2$ agrees with reports by \cite{zhang2024enhancing, datye2022strain, smithe2018high, radisavljevic2013mobility}, for MoSe$_2$ with \cite{zhang2021rapid, li2020sub, li2017scalable, chamlagain2014mobility}, and for WS$_2$ with \cite{yang2024biaxial, wang2021electron, ovchinnikov2014electrical} (see Fig.~\ref{fig:exp_validation}a). Similarly, for p-type TMDs, our results for MoSe$_2$ align with \cite{nutting2021electrical, chen2017highly, li2016isoelectronic}, for WSe$_2$ with \cite{ghosh2025high, movva2015high, pradhan2015hall, allain2014electron}, and for MoTe$_2$ with  \cite{bae2021mote2, shang2020situ, pradhan2014field} (see Fig.~\ref{fig:exp_validation}b). These comparisons, performed under realistic device conditions ($n = p = 10^{13}$\,cm$^{-2}$, $n_{\text{imp}} = 5\times10^{12}$\,cm$^{-2}$, SiO$_2$ dielectric environment), confirm the accuracy of our computational framework in capturing the baseline transport properties. The chosen parameters represent a central, experimentally relevant point within the broad range reported in the cited studies, ensuring our model's quantitative predictions are directly comparable to real device data.

Furthermore, our framework accurately captures the dielectric-dependent $\mu_\mathrm{e0}$ and $\mu_\mathrm{h0}$ at room temperature. For n-type MoS$_2$ (see Fig.~\ref{fig:exp_validation}c), our calculated mobility ranges across SiO$_2$, Al$_2$O$_3$, and HfO$_2$ dielectrics fully encompass the experimental values reported in \cite{chen2016chemical, smithe2016intrinsic, mahlouji2024influence, huang2023performance, xu2018effects}. Similarly, for p-type WSe$_2$ (see Fig.~\ref{fig:exp_validation}d), our predictions for SiO$_2$ and Al$_2$O$_3$ dielectrics show excellent agreement with experimental data from \cite{jang2022fermi, campbell2016field, pudasaini2017high, ghosh2025high}. These comparisons are performed under realistic device conditions ($n = p =  10^{11}$--$10^{13}$\,cm$^{-2}$, $n_{\text{imp}} = 10^{12}$--$10^{13}$\,cm$^{-2}$). This comprehensive validation across different dielectric environments reinforces the predictive capability of our computational framework for realistic device configurations.

\begin{figure}[htp]
\centering
\includegraphics[width=0.95\textwidth]{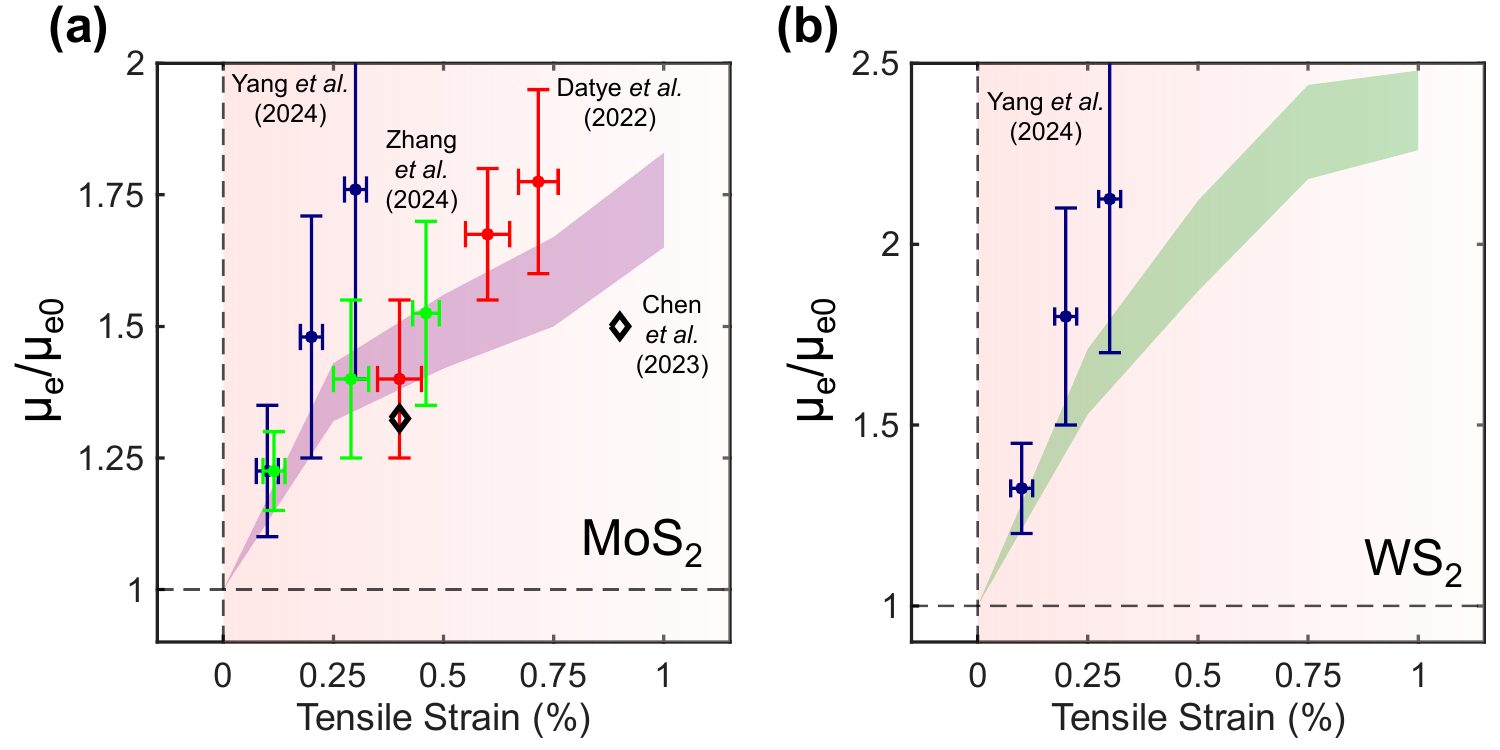}
\caption{\textbf{Experimental validation of strain-induced electron mobility enhancement in n-type TMDs.} 
\textbf{a,} Theoretical prediction and experimental validation of strain-induced electron mobility enhancement for monolayer MoS$_2$, showing excellent agreement with results from \cite{zhang2024enhancing, chen2023mobility, yang2024biaxial, datye2022strain}. 
\textbf{b,} Strain-induced electron mobility enhancement for monolayer WS$_2$, validated against experimental data from \cite{yang2024biaxial}. Simulations for both panels span experimentally relevant parameters: $T$ = 300 K, $n = 10^{11}$--$10^{13}$\,cm$^{-2}$, $n_{\text{imp}} = 10^{12}$--$10^{13}$\,cm$^{-2}$, and various dielectric environments (SiO$_2$, Al$_2$O$_3$, HfO$_2$) to calculate the enhancement range for MoS$_2$ (purple) and WS$_2$ (green).}
\label{fig:exp_validation_enhanc}
\end{figure}

The excellent agreement extends to strain-induced enhancement. Our predictions for n-type MoS$_2$ (see Fig.~\ref{fig:exp_validation_enhanc}a) under biaxial strain quantitatively match recent experimental measurements by \cite{zhang2024enhancing, chen2023mobility, yang2024biaxial, datye2022strain}, and for WS$_2$ (see Fig.~\ref{fig:exp_validation_enhanc}b) with data from \cite{yang2024biaxial}. This agreement is particularly significant as our model accurately reproduces enhancement trends at room temperature across different dielectric environments (SiO$_2$, Al$_2$O$_3$, HfO$_2$) and practical device conditions ($n = 10^{11}$--$10^{13}$\,cm$^{-2}$, $n_{\text{imp}} = 10^{12}$--$10^{13}$\,cm$^{-2}$) where previous theoretical studies employing the effective mass approximation \citep{phuc2018tuning, yu2015phase, hosseini2015strainmos2, hosseini2015strain, kumar2024strainfet} have not reported quantitatively predicting the enhancement observed in experiments \cite{yang2024biaxial, zhang2024enhancing, chen2023mobility, datye2022strain}.

For p-type TMDs, while direct experimental validation is currently limited due to the scarcity of systematic strain-dependent hole mobility theoretical studies, our computational predictions provide a compelling roadmap for future experimental investigations. The predicted enhancement trends for MoSe$_2$, WSe$_2$, and MoTe$_2$ under compressive strain are physically robust and consistent with the same mechanisms validated for n-type TMDs. We anticipate that our work will stimulate experimental efforts to explore strain engineering of hole transport in p-type TMDs, particularly given the superior performance predicted for WSe$_2$.

The fundamental advancement of our work lies in its complete physical treatment of carrier transport. Unlike prior studies that considered
individual scattering mechanisms or idealized conditions, our framework incorporates all relevant intrinsic and extrinsic scattering processes within a unified computational approach, thereby reinforcing confidence in our predictions of mobility and its enhancement due to strain. Our transport framework reveals that the strain-induced mobility enhancement primarily originates from the modification of IV scattering through the tuning of valley separations ($\Delta \mathrm{E}_{\mathrm{QK}}$ for electrons, $\Delta \mathrm{E}_{\mathrm{\Gamma Q}}$ for holes), while the baseline mobility value is determined through a combination of scattering mechanisms, both intrinsic and extrinsic. This physical insight explains why WS$_2$ shows the strongest electron mobility enhancement (278 meV/\%$\varepsilon$ in $\Delta \mathrm{E}_{\mathrm{QK}}$) and WSe$_2$ exhibits the best hole mobility improvement (341 meV/\%$\varepsilon$ in $\Delta \mathrm{E}_{\mathrm{\Gamma K}}$) among the materials studied.

Our analysis demonstrated that strain engineering remains effective even in the presence of strong extrinsic scattering, though the enhancement magnitude is somewhat reduced compared to the intrinsic case. This robustness stems from the fact that while CI and SO phonon scattering are less sensitive to strain, the suppression of IV scattering continues to provide significant benefits. The parametric studies further establish that $\frac{\partial(\mu_\mathrm{e}/\mu_\mathrm{e0})}{\partial\varepsilon}$ and $\frac{\partial(\mu_\mathrm{h}/\mu_\mathrm{h0})}{\partial\varepsilon}$ persist across temperature, carrier concentrations, impurity densities, and different dielectric environment variations, confirming its universal character.

The practical implications of our findings are substantial. The identified enhancement rates of 2.31/\%$\varepsilon$ for WS$_2$ (electron) and 2.37/\%$\varepsilon$ for WSe$_2$ (hole) under realistic device conditions indicate notable improvements in carrier mobility, highlighting the promise of strain engineering for future transistor applications. Supplementary Figs. 7 and 8 further suggest that larger strains, up to 5\%, could yield even greater enhancements, motivating experimental efforts toward innovative substrate engineering or heterostructure design to access such high-strain regimes. Our computational framework also opens pathways to explore the effects of strain on thermoelectric performance \citep{guo2016biaxial}, valley polarization \citep{guo2022valley}, and other emergent quantum phenomena \citep{liu2024first}.

\section*{Methods}\label{sec11}

To systematically investigate the strain-dependent carrier mobility in monolayer TMDs, we developed a comprehensive multi-scale computational framework. The framework integrates first-principles calculations of material properties with a quantum transport model. It enables a prediction of carrier mobility and its enhancement under strain. The workflow consists of three primary stages: (i) first-principles calculations of electronic and vibrational properties, (ii) computation of energy-dependent scattering rates for all relevant mechanisms and carrier transport, and (iii) calculation of the strain-induced carrier mobility enhancement.

\subsection*{First-principles calculations}

The electronic and vibrational properties of monolayer TMDs under biaxial strain were investigated using first-principles calculations based on Density Functional Theory (DFT) as implemented in the Quantum ESPRESSO suite \cite{giannozzi2017advanced}. The Perdew-Burke-Ernzerhof (PBE) exchange-correlation functional within the generalized gradient approximation (GGA) was employed \cite{perdew1996generalized}. Projector-Augmented Wave (PAW) pseudopotentials were employed to explicitly account for the following valence electron configurations: Mo ($4p^6 4d^5 5s^1$), W ($5p^6 5d^4 6s^2$), S ($3s^2 3p^4$), Se ($4s^2 4p^4$), and Te ($5s^2 5p^4$), ensuring accurate representation of both electronic and vibrational properties \cite{giustino2017electron}. A vacuum space of $\geq 18\,\text{\AA}$ was introduced perpendicular to the surface plane to suppress interactions between periodic slabs. The plane-wave kinetic energy cutoff was set to 80~Ry for the wavefunctions, with a charge density cutoff of 320~Ry. The Brillouin zone was sampled using a $16 \times 16 \times 1$ Monkhorst-Pack $k$-point grid for structural relaxations and electronic property calculations. All atomic positions and lattice vectors were fully relaxed until the total energy and interatomic forces were converged to within $10^{-6}$~eV/atom and $10^{-3}$~eV/\AA, respectively. Phonon dispersion relations and the energies of specific phonon modes were computed using density functional perturbation theory (DFPT) \cite{baroni2001phonons}. For these calculations, a denser $24 \times 24 \times 1$ $q$-point grid was used, and a Marzari-Vanderbilt cold smearing of 0.01~Ry was applied to ensure stable convergence. The obtained phonon energies at high-symmetry points were key inputs for evaluating electron-phonon scattering matrices in the subsequent mobility calculations.

\subsection*{Carrier transport}

In this study, transport coefficients were determined using a comprehensive full band structure approach, surpassing the limitations of the effective mass approximation (EMA). While EMA treats the energy dispersion as parabolic near the band edges, providing analytical convenience, it fails to capture the non-parabolic, anisotropic, and multi-valley nature of the electronic structure in TMDs, particularly under strain. Such effects are central to transport behavior, as carriers in TMDs occupy multiple valleys (K and Q for electrons, and $\Gamma$ and K for holes), and strain notably modulates the band curvature and shifts valley energies. These features are not easily represented by an effective mass alone. By directly incorporating the first-principles band structure, $E({\bf{k}})$, our approach accurately computes energy-dependent quantities such as the group velocity, ${\bf{v}}(E)$, and the density of states, $D(E)$. Thus, the calculated scattering rates and mobility capture the details of the electronic structure, providing a physically robust and quantitatively reliable picture of charge transport in strained 2D materials. A detailed analysis is provided in Supplementary Information, Section S2.

Within the linear response regime, the carrier mobility can be obtained via the Kubo-Greenwood expression~\cite{lundstrom2002fundamentals}, which requires the energy-dependent momentum relaxation time $\tau_\mathrm{m}(E)$ as its key input. The momentum relaxation time, obtained through the Fermi's golden rule (FGR), connects the physics of microscopic scattering processes to macroscopic transport coefficients, which can be measured in experiments. Supplementary Fig. 1 represents the schematic representation of carrier scattering mechanisms in 2D TMDs. FGR provides the transition probability of an electron in eigenstate $\bf{p}$ to scatter into another eigenstate $\bf{p'}$ of the pure material as governed by the scattering potential~\cite{lundstrom2002fundamentals} and is given by Supplementary equation (S.1).

The total momentum relaxation time $\tau_\mathrm{m}(E)$ that enters the mobility calculation is governed by Matthiessen's rule, combining all independent scattering mechanisms \cite{lundstrom2002fundamentals}:
\begin{equation}
\label{eq:TotalRate_main}
\frac{1}{\tau_\mathrm{m}(E)} = \sum_{i} \frac{1}{\tau_{\mathrm{m},i}(E)}.
\end{equation}
Here, the summation includes contributions from ADP scattering (Supplementary equation (S.6)), ODP scattering (Supplementary equation (S.10)), POP scattering (Supplementary equation (S.13)), IV scattering (Supplementary equation (S.14)), PZ scattering (Supplementary equation (S.16)), CI scattering (Supplementary equation (S.19)), and SOP scattering from the dielectric environment (Supplementary equation (S.21)). The essential physical parameters that determine the strength of these scattering mechanisms, including deformation potentials, elastic constants, phonon energies, sound velocity, and mass density, were obtained entirely from our first-principles calculations. A complete list of these calculated parameters is provided in Supplementary Tables 1 and 2. DFT calculations also provide information on ${\bf{v}}(E)$ and $D(E)$, which combined with $\tau_\mathrm{m}(E)$, yields the near-equilibrium mobility: 
\begin{equation}
\label{eq:final_mu_main}
\mu = \frac{q}{2n} \int v^2(E) \, \tau_\mathrm{m}(E) \, D(E) \left( -\frac{\partial f_0}{\partial E} \right) dE,
\end{equation}
where $q$ is the elementary charge, while $n$ represents the electron density for n-type TMDs and must be replaced with the hole density, $p$, for the calculation of hole mobility in p-type TMDs. $f_0(E)$ is the equilibrium Fermi-Dirac distribution. Our approach to estimating carrier mobility connects the details of electronic structure and scattering processes to the macroscopic observables. All parameters, including $\tau_\mathrm{m}(E)$, $D(E)$, and ${{v}}(E)$, were calculated at each value of applied strain for all the TMDs under study.

\subsection*{Strain-induced carrier mobility tuning}

The biaxial strain magnitude, $\varepsilon$ was defined as $\varepsilon = \frac{a - a_0}{a_0} \times 100\%$, where $a_0$ is the optimized equilibrium lattice constant. The values of $a_0$ for all five TMDs considered are provided in Supplementary Table~1. Biaxial strain was applied by uniformly scaling the in-plane lattice vectors from $-1\%$ (compressive) to $+1\%$ (tensile) in $0.25\%$ increments. By systematically incorporating strain-induced changes in both the electronic band structure and phonon spectra, we quantitatively analyzed their individual and combined impacts on scattering times and carrier mobility. The key outcome is the normalized mobility enhancement, \(\mu/\mu_0\), and its strain sensitivity, \(\frac{\partial(\mu/\mu_0)}{\partial \varepsilon}\), where \(\mu_0\) is the mobility of the unstrained TMD. This strain-dependent metric highlights the fundamental mechanisms driving electron and hole mobility improvements in 2D TMDs under biaxial strain.

\section*{Supplementary information}

\section*{Acknowledgements}

This work was supported by the National Science Foundation (NSF) through the University of Illinois Urbana–Champaign Materials Research Science and Engineering Center under Award DMR-2309037. The authors acknowledge the use of facilities and instrumentation supported by NSF through the University of Illinois Materials Research Science and Engineering Center DMR-2309037. The authors also acknowledge partial support by the NSF through the Center for Advanced Semiconductor Chips with Accelerated Performance Industry-University Cooperative Research Center under NSF Cooperative Agreement No. EEC-2231625.

\section*{Author contributions}

S.M.T.S.A. performed all calculations, analyzed the results, and wrote the manuscript. H.L.Z. helped interpret the results and contributed to the manuscript. A.M.v.d.Z. guided the direction of the research, provided critical feedback, and contributed to the manuscript. S.R. supervised the entire project, advised on the research strategy, provided scientific guidance, and contributed to the manuscript.

\section*{Data availability}

The primary data supporting the conclusions of this work are included in the main text and the Supplementary Information. Any additional data can be obtained from the authors upon reasonable request.

\section*{Declarations}
The authors declare no competing financial interest.

\bibliography{sn-bibliography}

\end{document}


\renewcommand{\thepage}{S\arabic{page}}
\maketitle

\section{Scattering Rates}

The scattering of carriers is described by the transition probability per unit time given by Fermi's golden rule for transition from an initial state $\bf{p}$ to a final state $\bf{p}'$~\cite{lundstrom2002fundamentals}:
\begin{equation}
\label{eq:FGR}
S({\bf{p}}, {\bf{p}'}) = \frac{2\pi}{\hbar} \left| H_{\bf{p'p}} \right|^2 \delta_{\mathbf{p'}, \mathbf{p} \pm\hbar\mathbf{\beta}} \delta(E({\bf{p}'}) - E({\bf{p}}) - \Delta E),
\end{equation}
where $H_{\bf{p'p}}$ denotes the matrix element of the scattering potential, $U_\mathrm{s}(\mathbf{r})$, and is given as 
\begin{equation}
\label{eq:MatrixElement}
H_{\mathbf{p}'\mathbf{p}} = I({\bf{p}}, {\bf{p'}})\times 
\frac{1}{A}\int_A e^{-i\mathbf{p'} \cdot \mathbf{r}/\hbar} U_\mathrm{s}(\mathbf{r}) e^{i\mathbf{p} \cdot \mathbf{r}/\hbar} \mathrm{d}^2 \mathbf{r}.
\end{equation}
The overlap integral, $I(\mathbf{p}, \mathbf{p'})$, representing the overlap of the Bloch functions over the unit cell, is obtained numerically from the DFT calculations. 
The momentum conservation due to phonon scattering is reflected in the term $\delta_{\mathbf{p'}, \mathbf{p} \pm\hbar\mathbf{\beta}}$, where $\beta$ is the phonon wave-vector, while the term, $\delta(E({\bf{p}'}) - E({\bf{p}}) - \Delta E)$, captures the energy conservation of the scattering process with $\Delta E = \pm \hbar \omega_\beta $ depicting the energy change during the inelastic scattering process ($\hbar \omega_\beta$ is the phonon energy).

\begin{figure}[h!]
    \centering
    \includegraphics[width=0.4\textwidth]{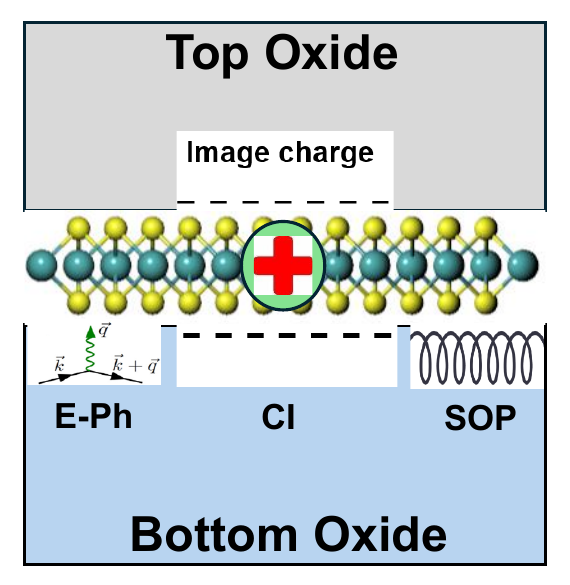}  
\caption{\textbf{Schematic representation of carrier scattering mechanisms in 2D TMDs.} Illustration of key scattering processes influencing charge transport in monolayer TMDs, including electron–phonon scattering due to lattice vibrations, Coulomb scattering from charged impurities in the substrate or the material, and SOP scatterings arising from interactions with polar optical modes of the oxide layers. }

    \label{fig:scattering_mechanisms}
\end{figure}

The momentum relaxation rate for a carrier with momentum $\bf{p}$ is defined by summing over all final states with the same spin, weighted by the efficiency of momentum randomization in each scattering event~\cite{lundstrom2002fundamentals}:
\begin{equation}
\label{eq:RelaxationRate}
\frac{1}{\tau_\mathrm{m} ({\bf{p}})} = \sum_{{\bf{p'}},\uparrow} S({\bf{p}}, {\bf{p'}}) \left(1- \frac{{\bf{p'}}\cdot {\bf{p}}}{p^2}\right) =  
\sum_{{\bf{p'}},\uparrow} S({\bf{p}}, {\bf{p'}}) \left(1- \frac{p'}{p}\cos \alpha \right),
\end{equation}
where $\alpha$ is the angle between the initial and final momentum vectors $\bf{p}$ and $\bf{p}'$. Supplementary Fig. \ref{fig:scattering_mechanisms} shows the schematic illustration of carrier scattering mechanisms in 2D TMDs. 
The total momentum relaxation rate is obtained by summing the contributions from acoustic deformation potential (ADP), optical deformation potential (ODP), polar optical phonon (POP), inter-valley (IV), piezoelectric (PZ), charged impurity (CI), and surface optical phonon (SOP) scattering according to
\begin{equation}
\label{eq:TotalRate}
\frac{1}{\tau_{\mathrm{m, total}} ({\bf{p}})} = \sum_{i} \frac{1}{\tau_{\mathrm{m}, i} ({\bf{p}})}.
\end{equation}
Note that as a result of the small momentum change in intra-valley scattering, the overlap integral is close to unity, as verified by our first-principles simulations, and is thus not explicitly included in $H_\mathrm{\mathbf{p'}\mathbf{p}}$.

\subsection{Acoustic Deformation Potential (ADP) Scattering}
 
The matrix element for ADP scattering is given as 
\begin{equation}
|H (\boldsymbol{\beta})|^2 = \frac{D_\mathrm{A}^2\beta^2 \hbar}{2\rho A \omega_\beta} \left( n_\mathrm{\omega} + \frac{1}{2} \mp \frac{1}{2} \right),     
\end{equation}
where $\rho$ is the mass density, $A$ is the area of the unit cell, $D_\mathrm{A}$ is the deformation potential constant ($D_\mathrm{Ae}$ for electron, $D_\mathrm{Ah}$ for hole), and $n_\omega = [e^{\hbar \omega_\beta/k_\mathrm{B}T_\mathrm{L}}-1]^{-1}$ is the Bose-Einstein distribution for phonons ($\hbar \omega_\beta$ is the acoustic phonon energy). The minus sign in the equation corresponds to phonon absorption, while the plus sign corresponds to phonon emission. Using Fermi's golden rule and using equation~(\ref{eq:RelaxationRate}) with linear phonon dispersion ($\hbar \omega_\beta = \beta v_\mathrm{s}$ with $v_\mathrm{s}$ being the sound velocity), the total momentum  relaxation rate due to absorption and emission is found to be 
\begin{equation}
\label{eq:tau_LA_final}
\frac{1}{\tau_\mathrm{m,ADP} (E)} = \frac{8D_\mathrm{A}^2}{\hbar^3 C_\mathrm{2D} v^2(E)}  \left[k_\mathrm{B}T_\mathrm{L}E + \frac{2v_\mathrm{s}E^2}{v(E)} \right],
\end{equation}
where $C_\mathrm{2D}$ is the material's elastic constant, and $v(E)$ is the group velocity. The material-specific parameters in equation (\ref{eq:tau_LA_final}) are determined from first-principles calculations. The 2D elastic constant, $C_{\mathrm{2D}}$, is defined as the second derivative of the total energy with respect to the in-plane strain
\begin{equation}
\label{eq:c2d}
C_{\mathrm{2D}} = \frac{1}{A_\mathrm{unit}} \frac{\partial^2 E_{\mathrm{tot}}}{\partial \epsilon^2},
\end{equation}
where $E_{\mathrm{tot}}$ is the total energy of the system, 
$A_\mathrm{unit}$ is the area of the unit cell, and $\epsilon = \Delta l / l_0$ is the applied strain. The acoustic deformation potential constant $D_\mathrm{A}$ quantifies the linear shift of a band edge energy, $E_{\mathrm{edge}}$, with strain:
\begin{equation}
\label{eq:def_pot}
D_\mathrm{A} = \frac{\partial E_{\mathrm{edge}}}{\partial \epsilon}.
\end{equation}
Supplementary Table~\ref{tab:tmd_properties} summarizes the values of the material parameters for ADP scattering obtained for the unstrained material, while the dependence of material properties on strain is plotted in Supplementary Fig.~\ref{fig:ADP_strain}.

\begin{table}[h!]
\centering
\caption{Calculated lattice constant ($a_0$), unit cell area ($A_\mathrm{unit}$), acoustic deformation potential constants for electrons ($D_\mathrm{Ae}$) and holes ($D_\mathrm{Ah}$) at the band edges, elastic constant ($C_\mathrm{2D}$), mass density ($\rho$), and sound velocity ($v_\mathrm{s}$) for five TMDs. Values are reported for the unstrained material.}
\label{tab:tmd_properties}
\begin{tabular}{lccccccc}
\toprule
TMDs & $a_0$ & $A_\mathrm{unit}$ & $D_\mathrm{Ae}$ & $D_\mathrm{Ah}$ & $C_\mathrm{2D}$ & $\rho$ & $v_\mathrm{s}$ \\
              & ($\times 10^{-10}$ m)          & ($\times 10^{-20}$ m$^{2}$)   & (eV)          & (eV)          & (J/m$^2$)        & ($\times 10^{-6}$ kg/m$^2$) & (km/s) \\
\midrule
MoS\textsubscript{2}  & 3.190 & 8.82  & 8.12 & 2.61 & 124.17 & 3.11 & 6.32 \\
MoSe\textsubscript{2} & 3.321 & 9.55  & 6.51 & 2.05 & 112.93 & 4.63 & 4.94 \\
MoTe\textsubscript{2} & 3.549 & 10.91 & 5.25 & 1.34 & 101.87 & 5.54 & 4.29 \\
WS\textsubscript{2}   & 3.187 & 8.81  & 6.95 & 1.85 & 143.13 & 4.89 & 5.41 \\
WSe\textsubscript{2}  & 3.325 & 9.58  & 6.83 & 1.89 & 121.53 & 6.23 & 4.42 \\
\bottomrule
\end{tabular}
\end{table}

\begin{figure}[h!]
\centering
\includegraphics[width=0.99\textwidth]{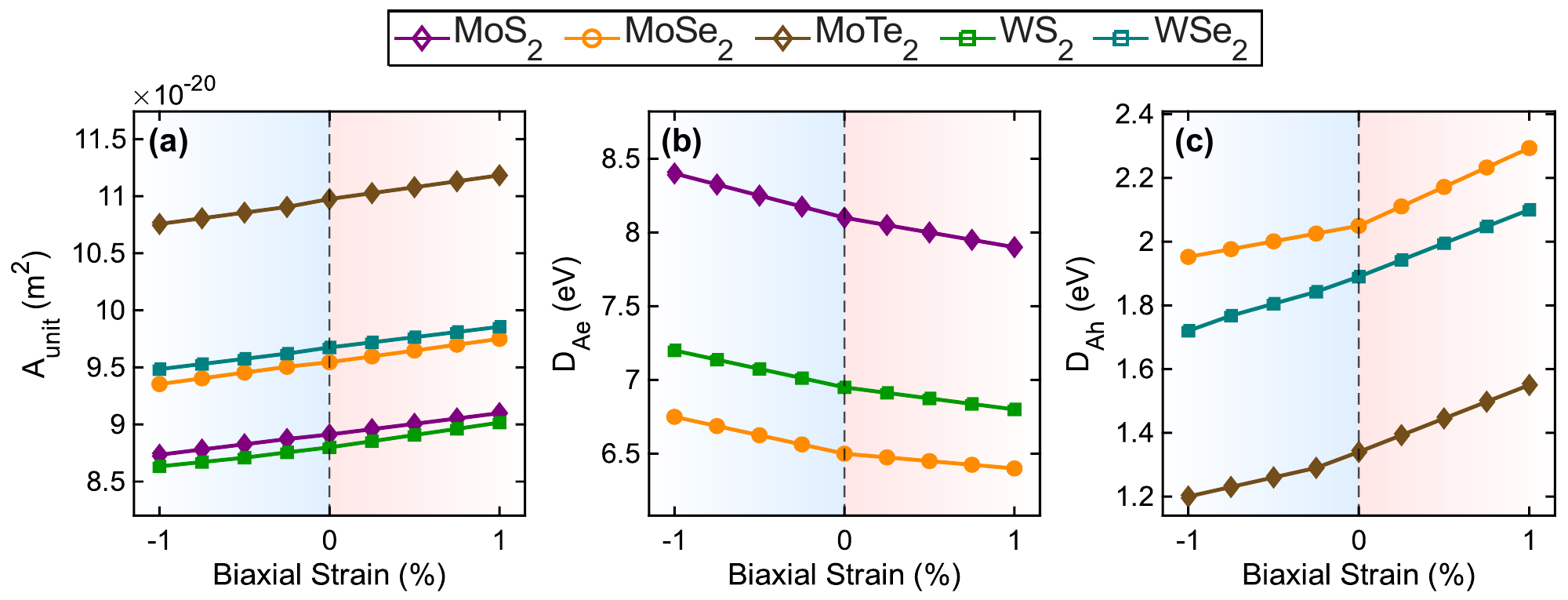}
\caption{\textbf{Strain dependence of material parameters governing ADP scattering.}
\textbf{a,} Unit cell area ($A_\mathrm{unit}$) as a function of biaxial strain.
\textbf{b,} Acoustic deformation potential for electrons ($D_\mathrm{Ae}$) showing systematic variation with applied biaxial strain for MoS$_2$ (purple), MoSe$_2$ (orange), and WS$_2$ (green). 
\textbf{c,} Acoustic deformation potential for holes ($D_\mathrm{Ah}$) of MoSe$_2$ (orange), WSe$_2$ (teal), and MoTe$_2$ (brown) exhibits the opposite trend compared to electrons.}
\label{fig:ADP_strain}
\end{figure}

Transverse acoustic (TA) phonon scattering is neglected in our calculations due to the significantly stronger intrinsic coupling between electrons and LA phonons via the deformation potential. Our first-principles results show that the LA deformation potential is consistently larger than that of the TA mode across all studied TMDs: MoS$_2$~(64\%), MoSe$_2$~(57\%), MoTe$_2$~(69\%), WS$_2$~(68\%), and WSe$_2$~(54\%). These findings are consistent with previous first-principles calculations by Kaasbjerg \textit{et al.}~\cite{kaasbjerg2013acoustic}, indicating that LA coupling is approximately 60\% stronger than TA in MoS$_2$. Furthermore, LA phonon dominance has also been noted qualitatively in prior literature~\cite{cheng2018limits}. Thus, TA phonons contribute minimally to intrinsic carrier scattering, justifying their exclusion from our mobility calculations.

\subsection{Optical Deformation Potential (ODP) Scattering}
The matrix element for ODP scattering is given as 
\begin{equation}
\label{eq:H_beta_sq_LO_1}
|H (\boldsymbol{\beta})|^2 = \frac{D_0^2 \hbar}{2\rho A \omega_0} \left( n_0 + \frac{1}{2} \mp \frac{1}{2} \right),
\end{equation}
where $D_0$ is the optical deformation potential, $\omega_0$ is the optical phonon frequency, and $n_0 = \left[\exp(\hbar \omega_0 / k_B T_L) - 1\right]^{-1}$ is the Bose-Einstein distribution function for optical phonons. In equation (\ref{eq:H_beta_sq_LO_1}), the minus sign corresponds to phonon absorption, while the plus sign corresponds to phonon emission. Using Fermi's golden rule and summing up the contributions of all final states with the same spin, the ODP momentum relaxation rate is found to be 
\begin{equation}
\label{eq:tau_LO_final}
\frac{1}{\tau_{\text{m, ODP}} (E)} = \frac{\pi D_0^2}{\rho \omega_0} \left( n_0 + \frac{1}{2} \mp \frac{1}{2} \right) D(E \pm \hbar \omega_0).
\end{equation}
This expression explicitly separates the contribution from phonon absorption, which requires final states at energy $E + \hbar \omega_0$, and phonon emission, which requires final states at energy $E - \hbar \omega_0$. The material-specific parameters for ODP scattering are obtained from first-principles calculations and listed in Supplementary Table~\ref{tab:optical}.

\begin{table}[h]
\centering
\caption{Calculated deformation potentials and phonon energies for intra-valley and inter-valley scattering. Values are reported for the unstrained material.}
\label{tab:optical}
\renewcommand{\arraystretch}{1.2}
\begin{threeparttable}
\resizebox{\textwidth}{!}{%
\begin{tabular}{
    >{\raggedright\arraybackslash}m{2cm}
    >{\centering\arraybackslash}m{1.5cm}
    >{\centering\arraybackslash}m{1.6cm}
    >{\centering\arraybackslash}m{1.6cm}
    >{\centering\arraybackslash}m{1.6cm}
    >{\centering\arraybackslash}m{1.6cm}
    >{\centering\arraybackslash}m{1.6cm}
    >{\centering\arraybackslash}m{2cm}
}
\toprule
\multirow{2}{*}{\textbf{Transition}} & 
\multirow{2}{*}{\textbf{}} & 
\multicolumn{5}{c}{\textbf{TMDs}} & 
\multirow{2}{*}{\textbf{Unit}} \\
\cmidrule(lr){3-7}
& & \textbf{MoS\textsubscript{2}} & \textbf{MoSe\textsubscript{2}} & \textbf{MoTe\textsubscript{2}} & \textbf{WS\textsubscript{2}} & \textbf{WSe\textsubscript{2}} & \\
\midrule
\multirow{2}{*}{K $\rightarrow$ K} 
& $D_{\mathrm{0}}$ & 2.08 & 2.18 & 3.97 & 1.09 & 0.83 & \SI{e8}{[eV/cm]} \\
& $\hbar\omega_{\mathrm{0}}$ & 46.53 & 33.15 & 30.24 & 45.96 & 30.04 & [meV] \\
\midrule
\multirow{2}{*}{Q $\rightarrow$ Q}  
& $D_{\mathrm{0}}$ & 4.85 & 4.17 & 4.96 & 2.34 & 1.87 & \SI{e8}{[eV/cm]} \\
& $\hbar\omega_{\mathrm{0}}$ & 48.34 & 34.46 & 31.23 & 47.02 & 31.12 & [meV] \\
\midrule
\multirow{2}{*}{$\Gamma \rightarrow \Gamma$}
& $D_{\mathrm{0}}$ & 3.48 & 3.82 & 4.97 & 1.48 & 2.23 & \SI{e8}{[eV/cm]} \\
& $\hbar\omega_{\mathrm{0}}$ & 50.13 & 36.34 & 32.86 & 48.93 & 33.09 & [meV] \\
\midrule
\multirow{3}{*}{K $\rightarrow$ Q}  
& $D_{\mathrm{if}}$ & 5.62 & 6.35 & 6.49 & 2.68 & 3.12 & \SI{e8}{[eV/cm]} \\
& $\hbar\omega_{\mathrm{if}}$ & 47.43 & 33.81 & 30.73 & 46.49 & 30.58 & [meV] \\
& $\Delta E_{\mathrm{fi}}$ & 115.14 & 77.89 & 195.19 & 108.07 & 65.56 & [meV] \\
\midrule
\multirow{3}{*}{$\Gamma \rightarrow$ K} 
& $D_{\mathrm{if}}$ & 5.98 & 5.19 & 6.67 & 2.87 & 4.89 & \SI{e8}{[eV/cm]} \\
& $\hbar\omega_{\mathrm{if}}$ & 48.33 & 34.74 & 31.55 & 47.44 & 31.56 & [meV] \\
& $\Delta E_{\mathrm{fi}}$ & 38.67 & 170.31 & 245.46 & 52.09 & 180.27 & [meV] \\
\bottomrule
\end{tabular}
}
\begin{tablenotes}
\footnotesize
\item[$\dagger$] Based on first-principles calculations, the phonon energies and deformation potentials associated with intra-valley and inter-valley scattering exhibit maximum deviations of $\pm$ 2\% and $\pm$ 4\%, respectively, from their unstrained values under applied biaxial strain.
\end{tablenotes}
\end{threeparttable}
\end{table}

\subsection{Polar Optical Phonon (POP) Scattering}
POP scattering represents a significant inelastic scattering mechanism in polar semiconductors like TMDs, where carriers interact with longitudinal optical (LO) phonons through the Fr\"ohlich interaction.
For the POP scattering, the matrix element is given as 
\begin{equation}
\label{eq:H_beta_sq_LO}
|H (\boldsymbol{\beta})|^2 =  \frac { \hbar q ^ { 2 } \omega _ { 0 } } { 2 \beta ^ { 2 } \kappa _ { 0 } \epsilon _ { 0 } A} \Bigl ( \frac { \kappa _ { 0 } } { \kappa _ { \infty } } - 1 \Bigr ) \left( n_{0} + \frac{1}{2} \mp \frac{1}{2} \right),
\end{equation}
where $q$ is the elementary charge, 
$\kappa_0$ ($\kappa_\infty$) is the low (high) frequency dielectric constant of the material, and $\epsilon_0$ is the free-space permittivity. The minus sign in the equation corresponds to phonon absorption, while the plus sign corresponds to phonon emission.

For the POP scattering, the momentum and energy conserving delta functions in equation (\ref{eq:FGR}) can be combined to yield
\begin{equation}
\label{eq:po_transition_rate}
S({\bf{p}}, {\bf{p}'}) = \frac{2\pi}{\hbar} |H (\boldsymbol{\beta})|^2  \frac{1}{\hbar v \beta}\delta\left( \pm \cos \theta + \frac{\hbar \beta}{2p} \mp \frac{\omega_{\beta}}{v \beta} \right),
\end{equation}
where $\theta$ is the angle between $\bf{p}$ and $\boldsymbol{\beta}$. Summing over the final momenta states that conserve spin gives 
\begin{multline}
\label{eq:tau_mp_refactored}
\frac{1}{\tau_{\mathrm{m,POP}}(E)} = 
\frac{q^2 \omega_{0} \left( \frac{\kappa_0}{\kappa_{\infty}} - 1 \right)}
     {8\pi \kappa_0 \epsilon_0 h v(E)} \Bigg\{ 
n_0 \left[ \sqrt{\frac{E + h \omega_0}{E}} 
     - \frac{h \omega_0}{E} \ln\left( \sqrt{\frac{E}{h \omega_0}} + \sqrt{ \frac{E + h \omega_0}{h \omega_0} } \right) \right] \\
+ (n_0 + 1) \left[ \sqrt{\frac{E - h \omega_0}{E}} 
     + \frac{h \omega_0}{E} \ln\left( \sqrt{\frac{E - h \omega_0}{h \omega_0}} + \sqrt{ \frac{E}{h \omega_0} } \right) \right]
\Bigg\},
\end{multline}
where $v(E)$ is the energy-dependent group velocity. This comprehensive derivation ensures proper treatment of POP scattering, accounting for both phonon absorption and emission processes across the entire energy spectrum. The $\kappa_0$ and $\kappa_{\infty}$ values of the five TMDs are taken from Klinkert \textit{et al.} \cite{klinkert20202}. The phonon energy ($\hbar \omega_0$) for POP scattering for five TMDs is obtained from first-principles calculations and listed in Supplementary Table \ref{tab:optical}.

The Fröhlich interaction is a long-range coupling that exclusively involves LO phonons in polar materials. LO phonon scattering is included in our calculations due to the strong Fröhlich interaction in polar TMDs. This long-range coupling dominates at elevated temperatures and significantly limits carrier mobility, unlike transverse optical (TO) phonons that lack this coupling mechanism \cite{cheng2018limits, kaasbjerg2014hot}.  Therefore, the contribution of TO phonons to intrinsic carrier scattering is negligible, justifying their exclusion from our mobility calculations.

\subsection{Inter-valley (IV) Scattering}

IV scattering, which involves transitions between different valleys, is treated within a framework analogous to ODP scattering in equation (\ref{eq:tau_LO_final}). Thus, the 2D IV scattering rate is given as
\begin{equation}
\label{eq:tau_IV_general}
\frac{1}{\tau_\mathrm{m,IV}(E)} = \frac{\pi D_\mathrm{if}^2 Z_\mathrm{f}}{\rho \omega_\mathrm{if}} |I(\mathbf{p}, \mathbf{p}')|^2 \left( n_\mathrm{if} + \frac{1}{2} \mp \frac{1}{2} \right) D_\mathrm{f} (E \pm \hbar \omega_\mathrm{if} - \Delta E_\mathrm{fi}),
\end{equation}
where $D_\mathrm{if}$ is the IV deformation potential constant, $Z_\mathrm{f}$ is the number of final valleys available for scattering, $\omega_\mathrm{if}$ is the characteristic IV phonon frequency, and $n_\mathrm{if}$ is the Bose-Einstein distribution. The top sign corresponds to absorption, while the bottom sign corresponds to emission. Since the phonon wavevector is essentially fixed by the valley locations, the overlap integral, \( I(\mathbf{p}, \mathbf{p}') \), is found to be a constant with respect to the final momenta states and is pulled out of the summation in equation (\ref{eq:RelaxationRate}) and thus appears as a multiplicative factor in equation (\ref{eq:tau_IV_general}). 
The 2D density of states of the final valley is denoted by $D_\mathrm{f}$, and $\Delta E_\mathrm{fi}$ represents the energy difference between the band edges (CBM for electrons, VBM for holes) of the final and initial valleys. For the TMDs discussed in this work, the valley degeneracies are 2 for the K valley, 6 for the Q valley, and 1 for the $\Gamma$ valley~\cite{kormanyos2018tunable}. The material-specific parameters for IV scattering, including the deformation potentials $D_\mathrm{if}$, phonon energies $\hbar\omega_\mathrm{if}$, and energy offsets $\Delta E_\mathrm{fi}$, are obtained from first-principles calculations and listed in Supplementary Table \ref{tab:optical}. The strain dependence of the overlap integral $|I(\mathbf{p}, \mathbf{p}')|^2$, which influences the IV scattering rate, is presented in Supplementary Fig.~\ref{fig:overlap}, as obtained from first-principles calculations.

\begin{figure}[h!]
\centering
\includegraphics[width=0.93\textwidth]{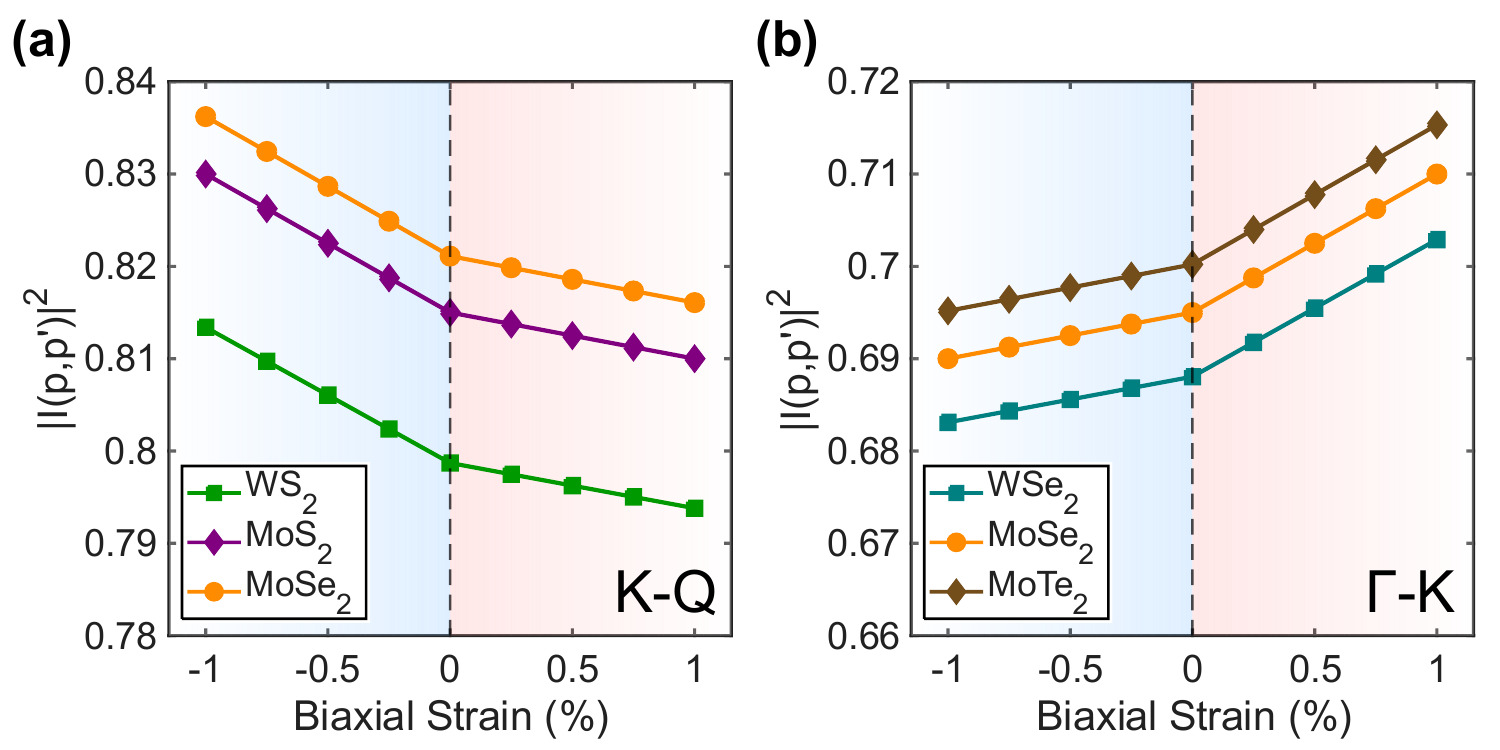}
\caption{\textbf{Strain dependence of the squared overlap integral $|I(\mathbf{p}, \mathbf{p}')|^2$ for inter-valley scattering in n-type and p-type TMDs.} 
\textbf{a,} $|I(\mathbf{p}, \mathbf{p}')|^2$ as a function of biaxial strain for three representative n-type TMDs: MoS$_2$ (purple), MoSe$_2$ (orange), and WS$_2$ (green), considering K--Q IV scattering. A clear trend is observed wherein $|I(\mathbf{p}, \mathbf{p}')|^2$ decreases with increasing tensile strain, while compressive strain leads to an enhancement in $|I(\mathbf{p}, \mathbf{p}')|^2$. 
\textbf{b,} Strain dependence of $|I(\mathbf{p}, \mathbf{p}')|^2$ for three p-type TMDs: MoSe$_2$ (orange), WSe$_2$ (teal), and MoTe$_2$ (brown), corresponding to $\Gamma$--K IV scattering. In contrast to the n-type case, tensile strain results in an increase in $|I(\mathbf{p}, \mathbf{p}')|^2$, whereas compressive strain leads to a reduction. The plotted values correspond to the direct scattering process between the specified valley momenta: for the K--Q transition, $\mathbf{p}=\mathbf{K}$ and $\mathbf{p}'=\mathbf{Q}$; for the $\Gamma$--K transition, $\mathbf{p}=\mathbf{\Gamma}$ and $\mathbf{p}'=\mathbf{K}$. 
}
\label{fig:overlap}
\end{figure}

\subsection{Piezoelectric (PZ) Scattering}

In polar semiconductors, the propagation of acoustic phonons can induce an electrostatic disturbance, a phenomenon termed piezoelectric scattering. The scattering rate attributed to this mechanism is substantially lower than that of polar optical phonon interactions. Nonetheless, its influence becomes non-negligible at low temperatures, where the population of optical phonons is significantly reduced. The matrix element for PZ scattering is given as 
\begin{equation}
|H (\boldsymbol{\beta})|^2 = \left(\frac{q e_{\mathrm{PZ}}}{\kappa_\mathrm{s} \epsilon_0}\right)^2 \frac{\hbar}{2\rho A \omega_\mathrm{s}} \left( n_{\omega} + \frac{1}{2} \mp \frac{1}{2} \right),
\end{equation}
where $q$ is the elementary charge, and $e_{\mathrm{PZ}}$ is the piezoelectric coefficient. 
The minus sign in the equation corresponds to phonon absorption, while the plus sign corresponds to phonon emission. Using Fermi's golden rule and summing up over all spin conserving final momenta states using equation~(\ref{eq:RelaxationRate}), we arrive at 
\begin{equation}
\label{eq:tau_pz_final}
\frac{1}{\tau_\mathrm{m, PZ} (E)} = \left( \frac{q e_{\mathrm{PZ}}}{\kappa_s \epsilon_0} \right)^2 
\frac{1}{2 \hbar C_\mathrm{2D}} 
\left[ \frac{k_\mathrm{B}T_\mathrm{L} v^2(E)}{E  (v^2(E)- v^2_\mathrm{s})} + 
\ln\left( \frac{v(E) + v_\mathrm{s}}{v(E)- v_\mathrm{s}} \right)  \right].
\end{equation}
Equation~(\ref{eq:tau_pz_final}) captures the combined contribution of both absorption and emission processes. The $\kappa_\mathrm{s}$ and $e_\mathrm{PZ}$ values for the five TMDs are taken from Klinkert \textit{et al.} \cite{klinkert20202} and Duerloo \textit{et al.} \cite{duerloo2012intrinsic}, respectively. Other 
material-specific parameters for PZ scattering, obtained from first-principles calculations, are listed in Supplementary Table \ref{tab:tmd_properties}.

\subsection{Charged Impurity (CI) Scattering}

The scattering rate due to charged impurities in a 2D system is derived within the Brooks-Herring framework. The scattering potential for a single ionized impurity is modeled by the screened Coulomb potential in real space~\cite{lundstrom2002fundamentals}:
\begin{equation}
\label{eq:screened_potential}
U_s({\bf{r}}) = \frac{q^2}{4\pi \kappa_\mathrm{s} \epsilon_0} \frac{e^{-{r}/L_\mathrm{D}}}{r},
\end{equation}
where $L_\mathrm{D}$ is the Debye screening length, represented by $L_\mathrm{D} = \sqrt{\frac{\kappa_\mathrm{s} \epsilon_0 k_\mathrm{B} T_\mathrm{L}}{e^2 n}}$, where $n$ represents the electron density for n-type TMDs and must be replaced with the hole density, $p$, for the calculation of hole mobility in p-type TMDs. The squared matrix element for CI scattering is given as
\begin{equation}
\label{eq:squared_matrix}
|H (\boldsymbol{\beta})|^2 = \frac{q^4}{4 \kappa_\mathrm{s}^2 \epsilon_0^2 A^2 (\beta^2 + {1}/{L_\mathrm{D}^2})}.
\end{equation}
For a random distribution of $N_\mathrm{I}$ impurities per unit area, the 
impurity scattering rate is found to be
\begin{equation}
\label{eq:tau_imp_final}
\frac{1}{\tau_\mathrm{m} ({E})} = \frac{q^4 N_\mathrm{I}}{8 \hbar \kappa_\mathrm{s}^2 \epsilon_0^2 E }
\left(1 - \frac{1}{\sqrt{1 + \frac{16E^2 L_\mathrm{D}^2}{\hbar^2 v^2(E)}}}\right).
\end{equation}

\subsection{Surface Optical Phonon (SOP) Scattering}

The SOP scattering mechanism arises from the polar vibrational modes in the surrounding dielectric materials, which interact with charge carriers in the semiconductor channel via long-range Coulomb forces. The theoretical treatment of this interaction relies on the dielectric response model of the surrounding media.

For a structure where the top oxide (tox) and bottom oxide (box) layers are approximated as semi-infinite dielectrics, and assuming no coupling between surface phonons and plasmons in the 2D channel, the energy dispersion of the SOP modes can be determined analytically, as shown in~\cite{ong2013theory}. The squared frequencies of the SOP modes associated with the bottom and top oxides are given by \( \omega_{\text{so,box}}^2 = \frac{-b + \sqrt{b^2 - 4ac}}{2a} \) and \( \omega_{\text{so,tox}}^2 = \frac{-b - \sqrt{b^2 - 4ac}}{2a} \), respectively. The coefficients \( a \), \( b \), and \( c \) depend on the dielectric properties of the surrounding materials and are defined as \( a = \epsilon_{\text{tox}}^\infty + \epsilon_{\text{box}}^\infty \), \( b = -(\epsilon_{\text{tox}}^0 + \epsilon_{\text{box}}^\infty) \omega_{\text{TO,tox}}^2 - (\epsilon_{\text{box}}^0 + \epsilon_{\text{tox}}^\infty) \omega_{\text{TO,box}}^2 \), and \( c = (\epsilon_{\text{tox}}^0 + \epsilon_{\text{box}}^0) \omega_{\text{TO,tox}}^2 \omega_{\text{TO,box}}^2\). Here, \( \epsilon^0 \) and \( \epsilon^\infty \) represent the static and high-frequency dielectric constants, respectively, and \( \omega_{\text{TO}} \) denotes the transverse optical phonon frequency of the dielectric. The matrix element for the electron-SOP interaction at a specific dielectric interface $i$ is given by
\begin{equation}
\label{eq:so_matrix_element}
|H (\boldsymbol{\beta})|^2 = \frac{e^2\hbar \omega_\mathrm{so, i}}{2 \epsilon_0  \beta A} \left( \frac{1}{\epsilon_\mathrm{i}^\infty + \epsilon_\mathrm{j}(\omega_\mathrm{so, i})} 
- \frac{1}{\epsilon_\mathrm{i}^0 + \epsilon_\mathrm{j}(\omega_\mathrm{so, i})}  \right) \left( n_\mathrm{so, i} + \frac{1}{2} \mp \frac{1}{2} \right),
\end{equation}
where $n_\mathrm{so, i} = \left[\exp(\hbar\omega_\mathrm{so, i}/k_B T) - 1\right]^{-1}$ is the Bose-Einstein occupation number for the SOP mode. The upper and lower signs correspond to phonon absorption and emission, respectively. The index $j$ denotes the opposite oxide layer (e.g., if $i = \text{tox}$, then $j = \text{box}$ and vice-versa). 
The calculated SOP energies for various dielectrics are presented in Supplementary Table~\ref{tab:dielectric_params}.

\begin{table}[htbp]
\centering
\caption{Dielectric material parameters and calculated surface optical phonon energies.}
\label{tab:dielectric_params}
\begin{threeparttable}
\begin{tabular}{lcccc}
\toprule
\textbf{Dielectrics} & \textbf{SiO\textsubscript{2}} & \textbf{Al\textsubscript{2}O\textsubscript{3}} & \textbf{HfO\textsubscript{2}} & \textbf{Unit} \\
\midrule
$\varepsilon_{\mathrm{tox}}^{0}$ & 3.9\tnote{a} & 12.53\tnote{a} & 23\tnote{a} & -- \\
$\varepsilon_{\mathrm{tox}}^{\infty}$ & 2.5\tnote{a} & 3.2\tnote{a} & 5.03\tnote{a} & -- \\
$\hbar\omega_{\mathrm{TO,tox}}$ & 55.6\tnote{a} & 48.18\tnote{a} & 12.4\tnote{a} & \si{\milli\electronvolt} \\
$\hbar\omega_{\mathrm{so,tox}}$ (This work) & 69.35 & 83.87 & 21.34 & \si{\milli\electronvolt} \\
$\hbar\omega_{\mathrm{so,box}}$ (This work) & 69.35 & 54.18 & 61.09 & \si{\milli\electronvolt} \\
\bottomrule
\end{tabular}
\begin{tablenotes}
\footnotesize
\item[a] Data from Ref.~\cite{fischetti2001effective}.
\end{tablenotes}
\end{threeparttable}
\end{table}

The total SOP scattering rate is the sum of the contributions from both the top and bottom oxide interfaces:
\begin{multline}
\label{eq:so_total_final_refactored}
\frac{1}{\tau_{\mathrm{m}, \mathrm{SO}}(E)} = 
\sum_{i=\text{tox, box}} 
\bigg\{ 
  \frac{e^2\omega_{so,i}}{2 \epsilon_0 \hbar v(E)}  
  \left( \frac{1}{\epsilon_\mathrm{i}^\infty + \epsilon_\mathrm{j}(\omega_\mathrm{so, i})} 
- \frac{1}{\epsilon_\mathrm{i}^0 + \epsilon_\mathrm{j}(\omega_\mathrm{so, i})}  \right) \times \\
  \Bigg[
    n_{\mathrm{so},i} \left( 
      \sqrt{1 - \frac{E}{E + \hbar \omega_{\mathrm{so},i}}}
      + \frac{\sqrt{E}}{ \left( \sqrt{E} + \sqrt{E + \hbar \omega_{\mathrm{so},i}} \right)}
    \right) \\
    + (n_{\mathrm{so},i} + 1) \left(
      \sqrt{1 - \frac{E - \hbar \omega_{\mathrm{so},i}}{E}}
      + \frac{\sqrt{E - \hbar \omega_{\mathrm{so},i}}}
             { \left( \sqrt{E} + \sqrt{E - \hbar \omega_{\mathrm{so},i}} \right)}
    \right)
  \Bigg]
\bigg\}
\end{multline}
Equation (\ref{eq:so_total_final_refactored}) provides the expression used to compute the SOP-limited scattering and includes the contributions of absorption and emission processes.

\clearpage

\section{Full-Band Versus Effective Mass Approximation Transport Models}

We present a systematic comparison of our full-band modeling approach against the conventional Effective Mass Approximation (EMA). While computationally economical, the EMA fundamentally oversimplifies the complex band structure of the TMDs by assuming a parabolic energy dispersion. This simplification fails to capture essential physics such as band non-parabolicity, multi-valley contributions, and the detailed density of states, all of which are paramount for accurately predicting carrier transport, particularly under the application of strain.

\begin{figure}[h!]
\centering
\includegraphics[width=0.9\textwidth]{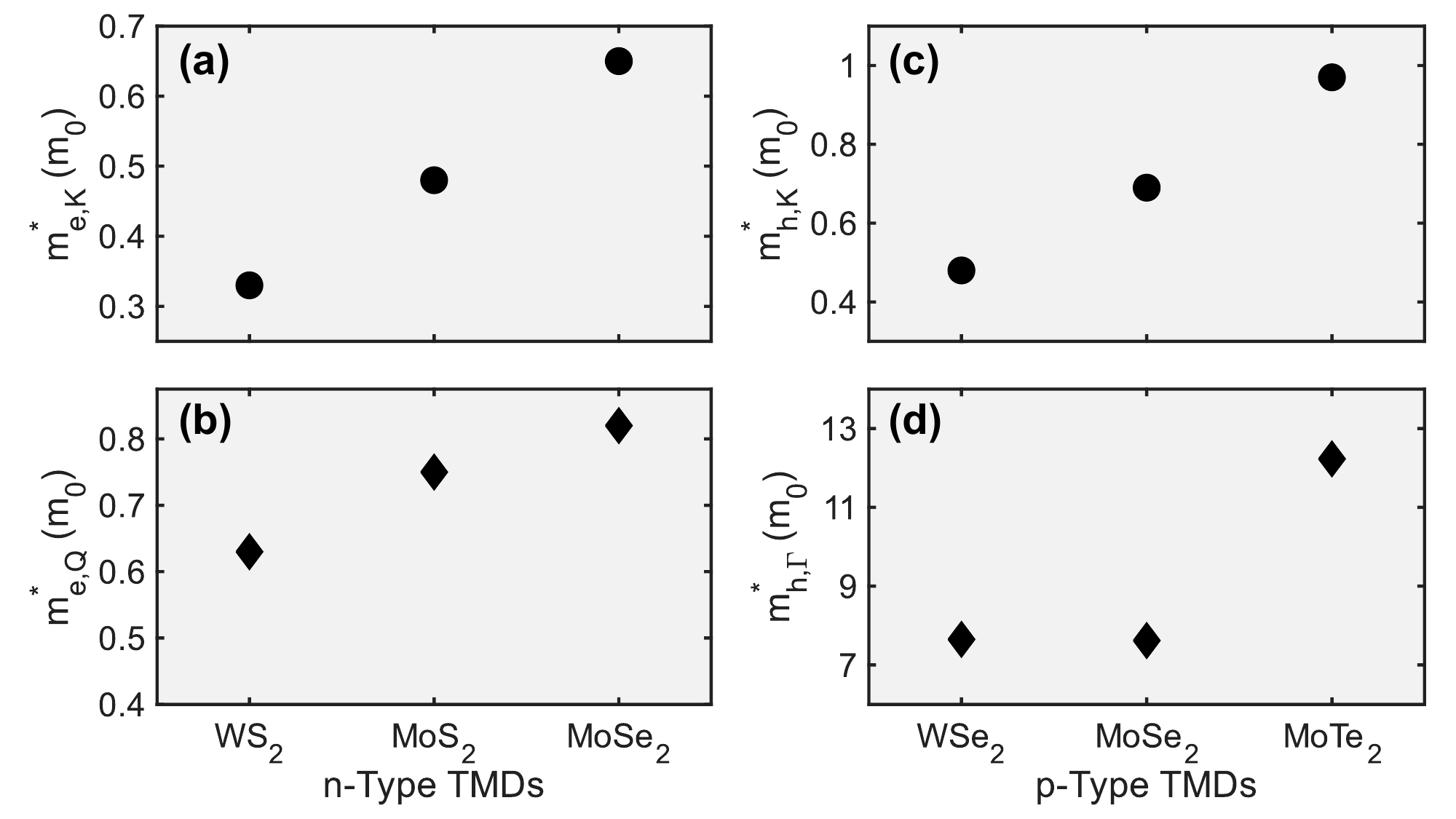}
\caption{\textbf{Carrier effective masses for n-type and p-type TMDs from first principles calculations.}
Electron effective mass $m_e^*$ for n-type monolayers at the \textbf{a,} K valley and \textbf{b,} Q valley.
Hole effective mass $m_h^*$ for p-type monolayers at the \textbf{c,} K valley and \textbf{d,} $\Gamma$ valley.
Results are shown for MoS$_2$, MoSe$_2$, WS$_2$ (n-type) and MoSe$_2$, WSe$_2$, MoTe$_2$ (p-type). WS$_2$ and WSe$_2$ exhibit the lowest electron and hole effective masses, respectively, indicating their superior intrinsic transport potential.}
\label{fig:effective_mass}
\end{figure}

EMA relies on an effective mass of carriers, obtained from the curvature of the electronic bands near the extrema points. Supplementary Figs. \ref{fig:effective_mass}a-d show the electron and hole effective masses for the relevant valleys as extracted from our first-principles electronic structure calculations. For n-type TMDs (Supplementary Figs.~\ref{fig:effective_mass}a and \ref{fig:effective_mass}b), WS$_2$ possesses the lowest electron effective mass at both the K and Q valleys, while MoSe$_2$ has the highest, a trend that predisposes WS$_2$ to higher mobility.
For p-type TMDs (Supplementary Figs.~\ref{fig:effective_mass}c and \ref{fig:effective_mass}d), WSe$_2$ exhibits the lowest hole effective mass, whereas MoTe$_2$ has the highest. Although these EMA-based masses provide an initial basis for performance trends (e.g., why WS$_2$ and WSe$_2$ are superior), they are insufficient for quantitative mobility prediction.

\begin{figure}[htbp]
\centering
\includegraphics[width=0.99\textwidth]{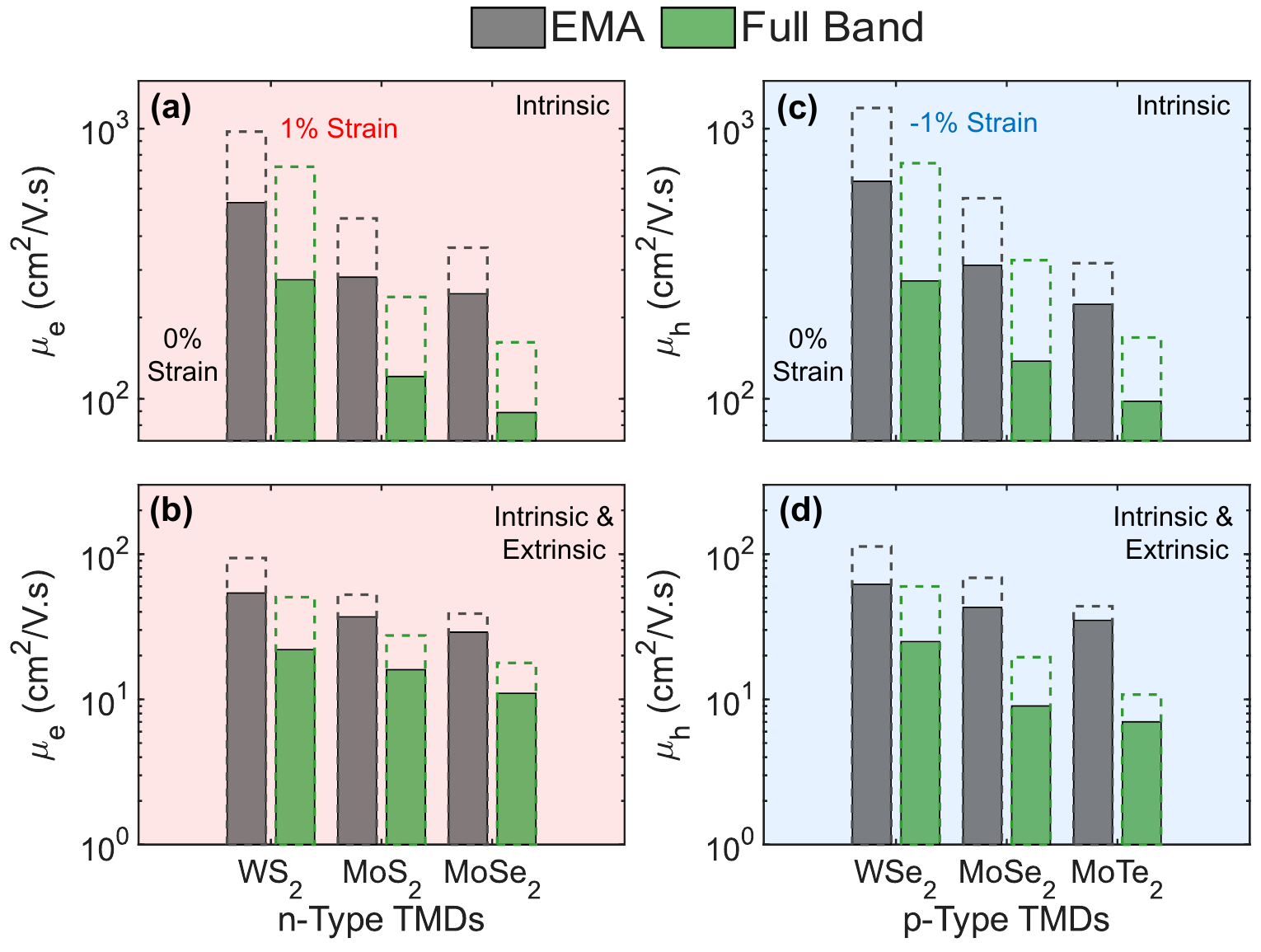} 
\caption{\textbf{EMA overestimates absolute carrier mobilities.} Calculated carrier mobilities for unstrained (solid bars) and strained (solid + patterned bars) cases: \textbf{a,} Intrinsic (ADP + ODP + POP + IV + PZ) electron mobility for n-type TMDs under 0\% and 1\% biaxial tensile strain. \textbf{b,} Electron mobility with extrinsic scattering (CI + SOP scattering). \textbf{c,} Intrinsic hole mobility for p-type TMDs under 0\% and -1\% biaxial compressive strain. \textbf{d,} Hole mobility with extrinsic scattering. The EMA (grey) systematically predicts overly optimistic absolute mobility values compared to the more realistic full-band model (green), highlighting the EMA's quantitative inaccuracy. Simulations performed at $T = 300$ K, $n = p = 10^{13}$ cm$^{-2}$, $n_{\mathrm{imp}} = 5 \times 10^{12}$ cm$^{-2}$, and a SiO$_2$ dielectric environment.}
\label{fig:mobility_comparison}
\end{figure}

The differences between the EMA and full-band transport models become pronounced when comparing predicted carrier mobilities. Supplementary Fig.~\ref{fig:mobility_comparison} shows the absolute mobilities for both unstrained and strained cases. A consistent and critical pattern emerges: the EMA systematically overestimates the absolute mobility compared to the more physically rigorous full-band model. This overestimation occurs because the EMA's parabolic band assumption artificially suppresses certain scattering mechanisms and misrepresents the density of states, leading to an overly optimistic projection of carrier mobility, both with and without strain.

The systematic underestimation of mobility enhancement by the EMA is quantitatively displayed in Supplementary Fig.~\ref{fig:enhancement_comparison}, which compares the predicted enhancement factors ($\mu/\mu_0$). For n-type WS\textsubscript{2} under tensile strain, the full-band model predicts an intrinsic enhancement factor of 2.63, notably exceeding the EMA prediction of 1.94 (see Supplementary Fig.~\ref{fig:enhancement_comparison}a). This discrepancy persists when extrinsic scattering is included, where the full-band mobility enhancement of 2.31 surpasses the EMA value of 1.76 by 31\% for WS$_2$ (see Supplementary Fig.~\ref{fig:enhancement_comparison}b). Similarly, for p-type WSe\textsubscript{2} under compressive strain, the full-band model yields a hole mobility enhancement of 2.71, substantially exceeding the EMA value of 2.16 (see Supplementary Fig.~\ref{fig:enhancement_comparison}c). Under extrinsic conditions, the full-band enhancement of 2.37 for WSe\textsubscript{2} remains 29\% higher than the EMA prediction of 1.83 (see Supplementary Fig.~\ref{fig:enhancement_comparison}d).

\begin{figure}[h!]
\centering
\includegraphics[width=0.99\textwidth]{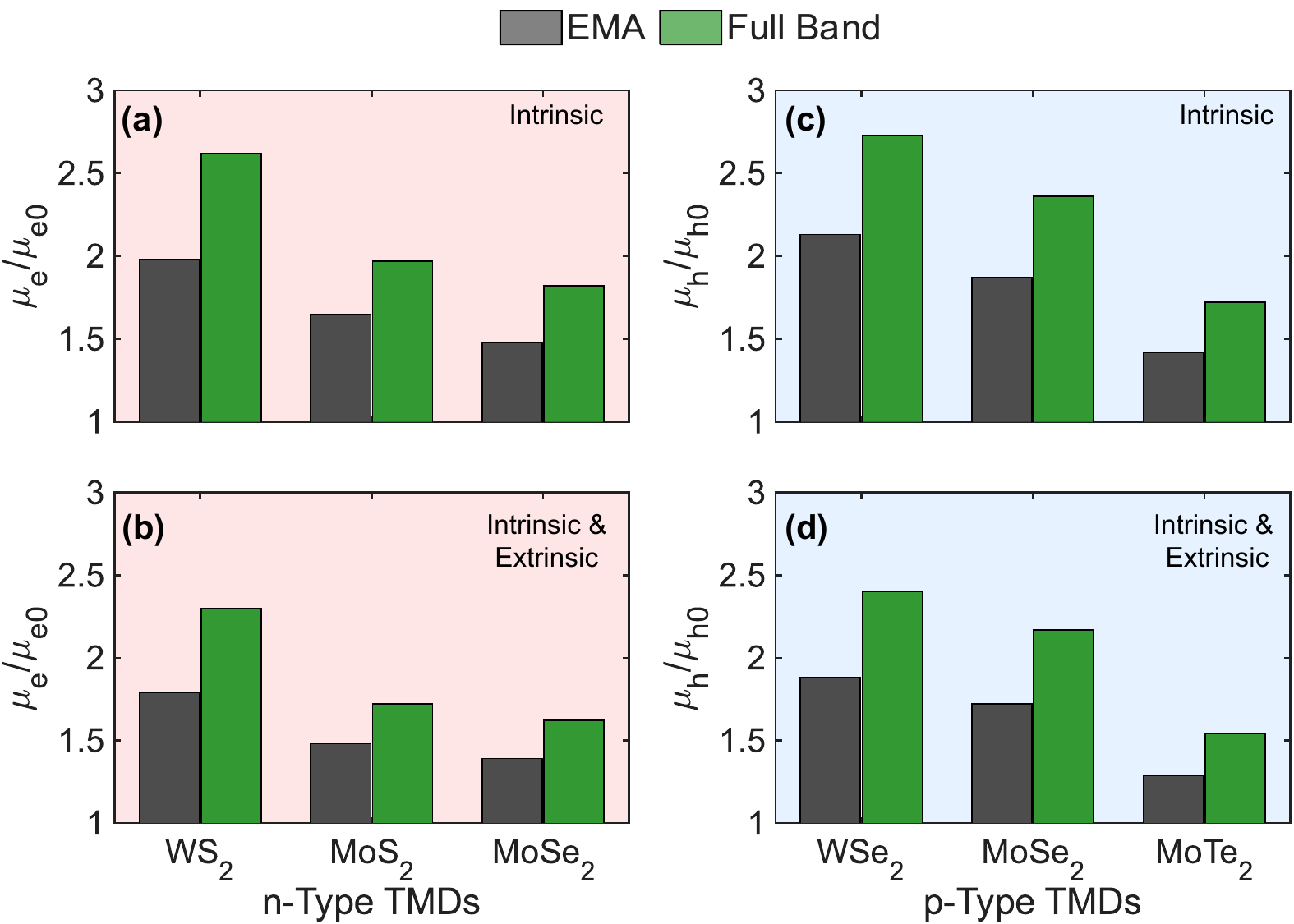} 
\caption{\textbf{Strain-induced carrier mobility enhancement factors from EMA and full-band models.} Electron mobility enhancement factor under 1\% biaxial tensile strain for n-type TMDs, calculated using the EMA (grey) and the full-band model (green) for \textbf{a,} intrinsic (ADP + ODP + POP + IV + PZ) and \textbf{b,} both intrinsic and extrinsic effects (CI + SOP scattering).
Hole mobility enhancement factor under -1\% biaxial compressive strain for p-type TMDs for \textbf{c,} intrinsic and \textbf{d,} both intrinsic and extrinsic effects.
The full-band model consistently predicts higher and more accurate enhancement factors across all materials and scattering regimes, demonstrating its critical role in quantitative strained-device prediction. Simulations performed at $T = 300$ K, $n = p = 10^{13}$\,cm$^{-2}$, $n_{\text{imp}} = 5 \times 10^{12}$\,cm$^{-2}$, and SiO$_2$ dielectric environment.}
\label{fig:enhancement_comparison}
\end{figure}

While the EMA model is unable to capture strain-induced band deformation beyond a simple effective mass change,
our full-band approach self-consistently computes the group velocity and density of states from the atomically-relaxed strained band structure, thereby accurately accounting for non-parabolicity and the shifting energetic contributions of different valleys to carrier mobility. Thus, a full-band approach offers a quantitatively precise measure of scattering rates,
ultimately yielding the higher, and experimentally verifiable, mobility values and enhancement factors. Consequently, the full-band model is not a mere incremental improvement but an indispensable tool for the predictive design of strain-engineered 2D TMD devices.

\section{High-Strain Carrier Mobility Enhancement (Up to 5\%)}

\begin{figure}[htp]
\centering
\includegraphics[width=0.95\textwidth]{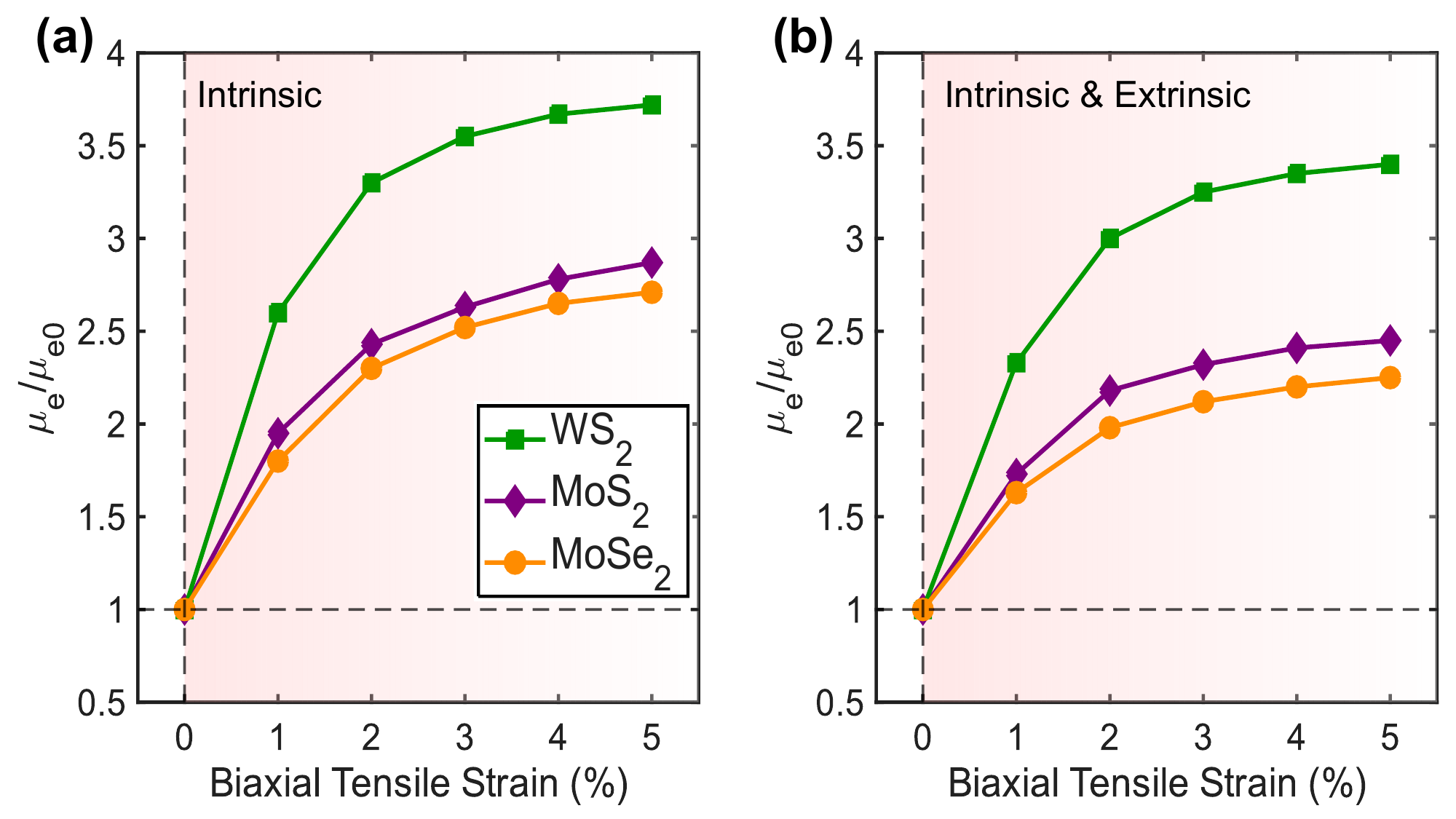}
\caption{\textbf{Enhanced electron mobility in n-type TMDs through tensile strain engineering (up to 5\%).} 
\textbf{a,} Intrinsic electron mobility enhancement considering ADP, ODP, POP, IV, and PZ scattering mechanisms under biaxial tensile strain. 
\textbf{b,} Total electron mobility enhancement incorporating both intrinsic and extrinsic effects (CI + SOP scattering) under large biaxial strain. Results are shown for MoS$_2$ (purple), MoSe$_2$ (orange), and WS$_2$ (green) at $T$ = 300 K, $n = 10^{13}$\,cm$^{-2}$, $n_{\text{imp}} = 5 \times 10^{12}$\,cm$^{-2}$, and SiO$_2$ dielectric environment. Tensile strain consistently enhances electron mobility even at high strain levels, though the enhancement rate diminishes compared to lower strain regimes due to reduced rates of $\Delta \mathrm{E}_{\mathrm{QK}}$ change. WS$_2$ maintains the most significant improvement across both scattering regimes, demonstrating the robustness of strain engineering even in larger deformation conditions. The corresponding unstrained electron mobilities ($\mu_\mathrm{e0}$) for each case are reported in the main text (see Fig. 2).}
\label{fig:s1_ntype_high_strain}
\end{figure}

\begin{figure}[htp]
\centering
\includegraphics[width=0.95\textwidth]{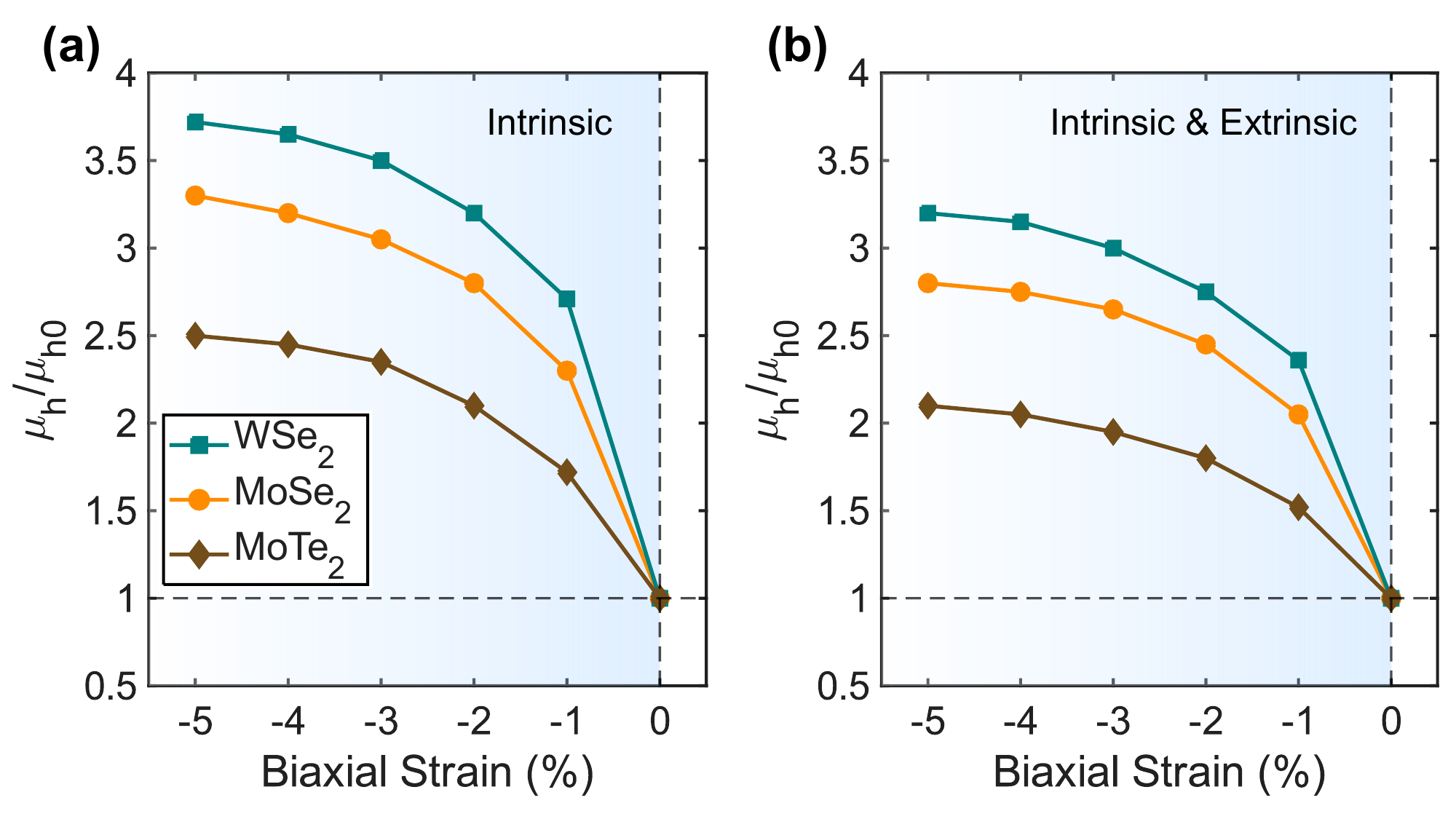}
\caption{\textbf{Enhanced hole mobility in p-type TMDs through compressive strain engineering (up to 5\%).} 
\textbf{a,} Intrinsic hole mobility enhancement considering ADP, ODP, POP, IV, and PZ scattering mechanisms under biaxial compressive strain. 
\textbf{b,} Total hole mobility enhancement incorporating both intrinsic and extrinsic effects (CI + SOP scattering) under large biaxial strain. Results are shown for MoSe$_2$ (orange), WSe$_2$ (teal), and MoTe$_2$ (brown) at $T$ = 300 K, $p = 10^{13}$\,cm$^{-2}$, $n_{\text{imp}} = 5 \times 10^{12}$\,cm$^{-2}$, and SiO$_2$ dielectric environment. Compressive strain consistently enhances hole mobility even at high strain levels, though the enhancement rate diminishes compared to lower strain regimes due to reduced rates of $\Delta \mathrm{E}_{\mathrm{\Gamma K}}$ change. WSe$_2$ maintains the most significant improvement across both scattering regimes. The corresponding unstrained hole mobilities ($\mu_\mathrm{h0}$) for each case are reported in the main text (see Fig. 4).}
\label{fig:s2_ptype_high_strain}
\end{figure}

\clearpage

\bibliographystyle{ieeetr} 
\bibliography{sn-bibliography} 